\documentstyle[aps,preprint]{revtex}
\tighten
\begin{document}
\draft
\title{\bf Critical Collapse of the Exchange Enhanced Spin Splitting in 2-D Systems}

\author{$^{1}$D.R.~Leadley, $^{2}$R.J.~Nicholas, $^{3}$J.J.~Harris and 
$^{4}$C.T.~Foxon}

\address{$^{1}$ Department of Physics, University of Warwick, Coventry, CV4~7AL,~UK\\
$^{2}$Department of Physics, Clarendon Laboratory, Parks Road, Oxford, OX1~3PU,~UK\\
$^{3}$Department of Electronic and Electrical Enginering, University College, London, 
WC1E~7JE,~UK\\
$^{4}$Department of Physics, Nottingham University, University Park, Nottingham, 
NG7~2RD,~UK}

\date{Submitted to Phys. Rev. B May 11, 1998}

\maketitle

\begin{abstract}
The critical filling factor $\nu_c$ where Shubnikov-de~Haas oscillations become spin split is 
investigated for a set of GaAs-GaAlAs heterojunctions.  Finite temperature magnetoresistance 
measurements are used to extract the value of $\nu_c$ at zero temperature.  The critically point 
is where the disorder potential has the same magnitude as the exchange energy, leading to the 
empirical relationship $\nu_c = g^*n_e\tau_sh/2m_0$.  This is valid for all the samples studied, 
where the density $n_e$ and single particle lifetime $\tau_s$ both vary by more than an order of 
magnitude and $g^*$ the exchange enhanced $g$-factor has a weak dependence on density.  For 
each sample the spin gap energy shows a {\em linear} increase with magnetic field.  
Experiments in tilted magnetic field show the spin gap is the {\em sum} of the bare Zeeman 
energy and an exchange term.  This explains why measurements of the enhanced 
$g$-factor from activation energy studies in perpendicular field and the coincidence method in 
tilted fields have previously disagreed.
\end{abstract}
\pacs{73.40.Hm, 73.20.Dx, 72.20.Jv}

\section{Introduction}

Quantisation of the Hall effect was first recognized in 1980 \cite{kvk} and it is understood that 
plateaux appear in the off-diagonal component of the magnetoresistance tensor $\rho_{xy}$ 
whenever the Fermi energy lies in a mobility gap of the electronic density of states.  In an ideal 
sample this would be the cyclotron gap $\hbar\omega_c$.  Associated with the plateaux are 
minima in the diagonal component which lead to the appearance of Shubnikov-de~Haas 
oscillations (SdHO) in $\rho_{xx}$.  SdHOs were observed in semiconductors as long ago as 
1966 \cite{stiles} and even in these earliest observations spin and valley splittings were seen to 
modify the underlying $1/B$ periodicity of the oscillations.  In particular the spin splitting 
appeared to be much stronger than expected from just a bare Zeeman gap of $g_0\mu_BB$, 
leading to the idea of an enhanced $g$-factor $g^*$ \cite{nich}.  This enhancement is due to 
many-body electron interactions that are introduced by forming a widely separated electron-hole 
pair with reversed spin. It is often referred to as {\em exchange enhancement} (after the 
exchange-correlation terms encountered in Hartree Fock calculations) and leads to the spin gap 
being written as:
\begin{equation}
\Delta_{spin}=g*\mu_BB=g_0\mu_BB+E_{ex}.
\end{equation}
Notice particularly that this is the {\em sum} of the bare Zeeman energy and an exchange 
energy $E_{ex}$ and that, unless $E_{ex}\propto B$, $g^*$ will itself be a function of field as 
opposed to a simple multiple of $g_0$.  Several semi-empirical versions of this equation have 
been adopted to model experimental results but it is the calculation of $E_{ex}$ that has 
particularly excerised theoreticians for the past thirty years \cite{au,kh,fs}.  

It is now timely to revisit the exchange enhanced spin splitting for two reasons. First, there is 
currently great interest in complex spin textures, or skyrmions, at low odd integer filling factors 
$\nu$ \cite{skyrmions}.  This work has shown a rich spectrum of spin excitations and serves to 
remind us how poorly even the basic spin splitting is understood.  The calculations indicate that 
at $\nu=1$ skyrmion-antiskyrmion pairs will always have a lower excitation energy than 
electron-hole pairs, but for higher filling factors the preferred excitation depends crucially on 
how the force laws are treated for real systems both for single spin excitations and the larger 
textures \cite{cooper}.  Secondly, there is a prediction by Fogler and Shklovskii (FS) \cite{fs} 
that the exchange enhancement may be destroyed by disorder and lead to the collapse of the spin 
splitting at a critical filling factor $\nu_c$.  FS gives expressions for $\nu_c$ in terms of sample 
parameters relevant to GaAs/GaAlAs heterojunctions which can be compared with experiments 
\cite{wong,shikler}.

These issues will be discussed further in the remainder of the paper, but first the phenomenon of 
spin splitting will be introduced with reference to some experimental data.  The behavior as a 
function of temperature and sample parameters will then be investigated, both through the filling 
factor where the spin split SdH maxima appear and by considering the size of the energy gap at 
odd integer $\nu$.  The temperature dependence of the separation between the maxima is found 
to scale onto a universal curve for all filling factors and all samples in a way that suggests a 
phase transition.  By understanding how $\nu_c$ varies with temperature we are able to extract 
the value at $T=0$ from the finite temperature data, allowing meaningful comparison with 
theory.  An empirical relationship between the critical filling factor and the measured sample 
parameters is established which justifies the idea of disorder driven collapse of the exchange 
enhancement.  By measuring the energy gap at different large odd filling factors within each 
sample we find a {\em linear} increase in $\Delta_{spin}$ with magnetic field.  Although a 
$\sqrt{B}$ increase would be expected for an exchange energy driven by the Coulomb energy, 
where correlation takes place over the magnetic length, this is not the case with multiple Landau 
levels (LL) occupied when the correlation length is set by the Fermi wavevector $k_F$ and a 
linear behavior results.
Finally, experiments in tilted field will be reported which show that the exchange part of 
$\Delta_{spin}$ depends only on the component of magnetic field perpendicular to the 
two-dimensional electron gas and that the increase in spin gap is due to the bare Zeeman term 
only.  This explains why $g$-factors obtained from the coincidence method in tilted fields do 
not agree with those found from activation studies in perpendicular field \cite{nich,usher}.

\section{The Spin Splitting Phenomenon }

The samples studied here are the well known GaAs/GaAlAs single heterojunctions grown at 
Philips Research Laboratories, Redhill having undoped spacer layer thickness in the range 100 
\AA$<L_z<$3200 \AA.  The samples cover a wide range of density and mobility and include 
examples with low disorder, which exhibit the fractional quantum Hall effect, as well as highly 
disordered samples, which have very wide integer quantum Hall plateaux.  Table~\ref{samples} 
lists some relevant sample parameters, measured at 1~K.  The experiments consist of 
magnetoresistance measurements performed over a range of temperatures from 50~mK to 4.2~K 
with the magnetic field normal to the sample, except in the final section where we consider tilted 
field measurements. The temperature $T$ is measured and stabilized using a ruthenium oxide 
resistor, mounted to be in thermal equilibrium with the sample and to have negligible 
magnetoresistance.

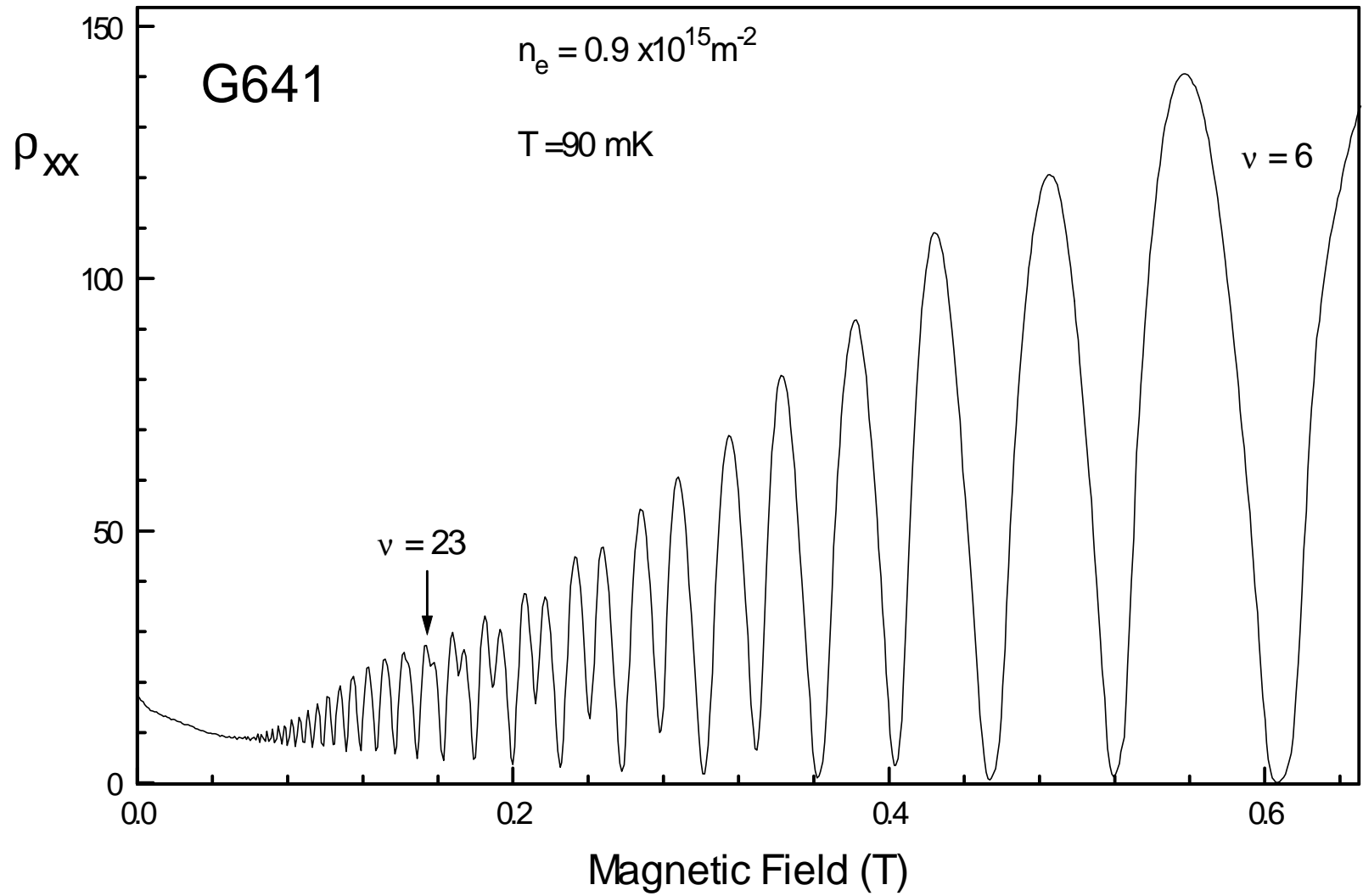
\begin{figure}
\caption{Magnetoresistance of sample G641 at 90 mK showing the spin splitting collapse as a 
function of magnetic field.}
\label{fig:sdh641}
\end{figure}

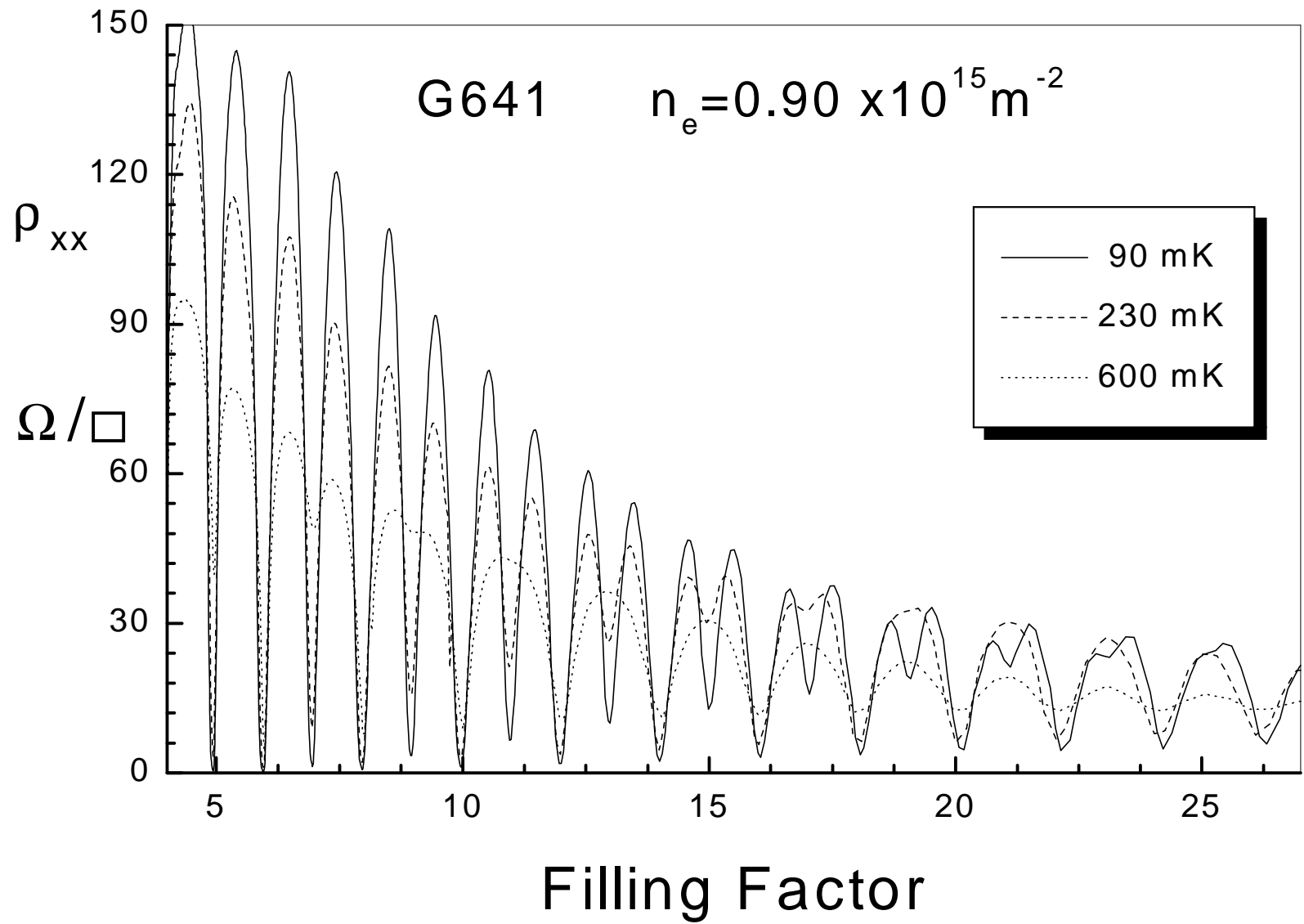
\begin{figure}
\caption{The data from Fig.~1 replotted as a function of filling factor together with data taken at 
230~mK and 600~mK which show how $\nu_c$ decreases at higher temperature.}
\label{fig:rhonu641}
\end{figure}

A typical low temperature recording of the diagonal resistivity $\rho_{xx}$ for a high mobility 
sample G641 is shown in Fig.~\ref{fig:sdh641}.  At high magnetic fields, in this case 
$B>0.5T$, the SdHOs are spin split and the maxima are evenly spaced, appearing at all 
half-integer filling factors.  At low field, $B<0.15T$ in Fig.~\ref{fig:sdh641}, the maxima are 
again evenly spaced but here there is no spin splitting so they appear at odd integer filling 
factors.  By examining either of these regions in isolation it would be impossible to tell whether 
or not the SdHOs were spin split without additional information.  In the intermediate region, 
whose extent depends on the temperature of the measurement, the spin splitting is partially 
resolved.  In this region, the spin split minima are less deep than those due to cyclotron splitting 
since their energy gaps are smaller.  Also towards lower magnetic fields the spin split maxima 
converge with a rapid change from a spacing $\delta\nu=|\nu_{\uparrow}-
\nu_{\downarrow}|=1$ at high fields to $\delta\nu=0$ where the spin splitting disappears.  In 
Fig.~\ref{fig:sdh641}, spin splitting can clearly be seen at $\nu=23$ and at higher magnification 
it can just be made out for the next two peaks but no more.  Clearly this is only a qualitative 
judgement of $\nu_{c}$ whereas a quantitative definition is required.  In FS this was taken to be 
the point where $\delta\nu=0.5$ at $T=0$.  Although the zero temperature requirement 
simplifies the theory, it adds complications when one only has experimental results at finite 
temperature as can be seen in Fig.~\ref{fig:rhonu641}.  Data from several temperatures are re-
plotted as a function of filling factor, allowing the convergence of the maxima to be seen more 
clearly.  While the last spin split peak is at $\nu=25$ in the 90~mK data, by 600~mK it appears 
at more than twice the magnetic field at $\nu=9$.  One aspect of this paper will be to establish a 
reliable way of extrapolating the real experimental data to the ideal $T=0$ situation.

Let us now examine some differences between the regular SdHOs and the spin splitting by again 
referring to the magnetic field dependence shown in Fig.~\ref{fig:sdh641}.  The SdHOs are 
damped exponentially towards low field according to the well know Lifshitz-Kosevich formula 
\cite{lk}:
\begin{equation}
\Delta\rho_{xx} \propto \frac{X}{\sinh{X}} \exp\left(-\frac{\pi}{\omega_c\tau_s}\right) 
\cos\left(\pi(\nu+1/2)\right),
\label{eqn:LK}
\end{equation}
where the factor containing $X=2\pi^2kT/E_g$ arises from the width of the Fermi function at 
finite temperature.  When there is no spin splitting the gap $E_g$ is the cyclotron energy 
$\hbar\omega_c$.  Eq.~(\ref{eqn:LK}) shows that the damping at $T=0$ is just determined by 
the single particle scattering time $\tau_s$, but is larger at higher temperatures when the Fermi 
function becomes smeared. In practice this means that more SdHOs will be seen if the 
experiments are performed at lower temperatures and with higher resolution. Thus there is not a 
maximum filling factor $\nu_c^{SdH}$ at zero temperature and even when the LL broadening 
is significantly larger than $\hbar\omega_c$ small oscillations will still occur in the 
conductivity. However, the exponential damping will set a limit in a real experiment where the 
last oscillation observed is determined by experimental noise. Practically, great care also has to 
be taken to sweep the magnetic field sufficiently slowly and to obtain enough data points per 
oscillation.  In our imperfect noisy experiments, SdHOs have been regularly observed at filling 
factors in excess of 100 and in the best cases with $\nu>150$. Values of $\tau_s$ deduced from 
Eq.~(\ref{eqn:LK}) are included in Table~\ref{samples}.

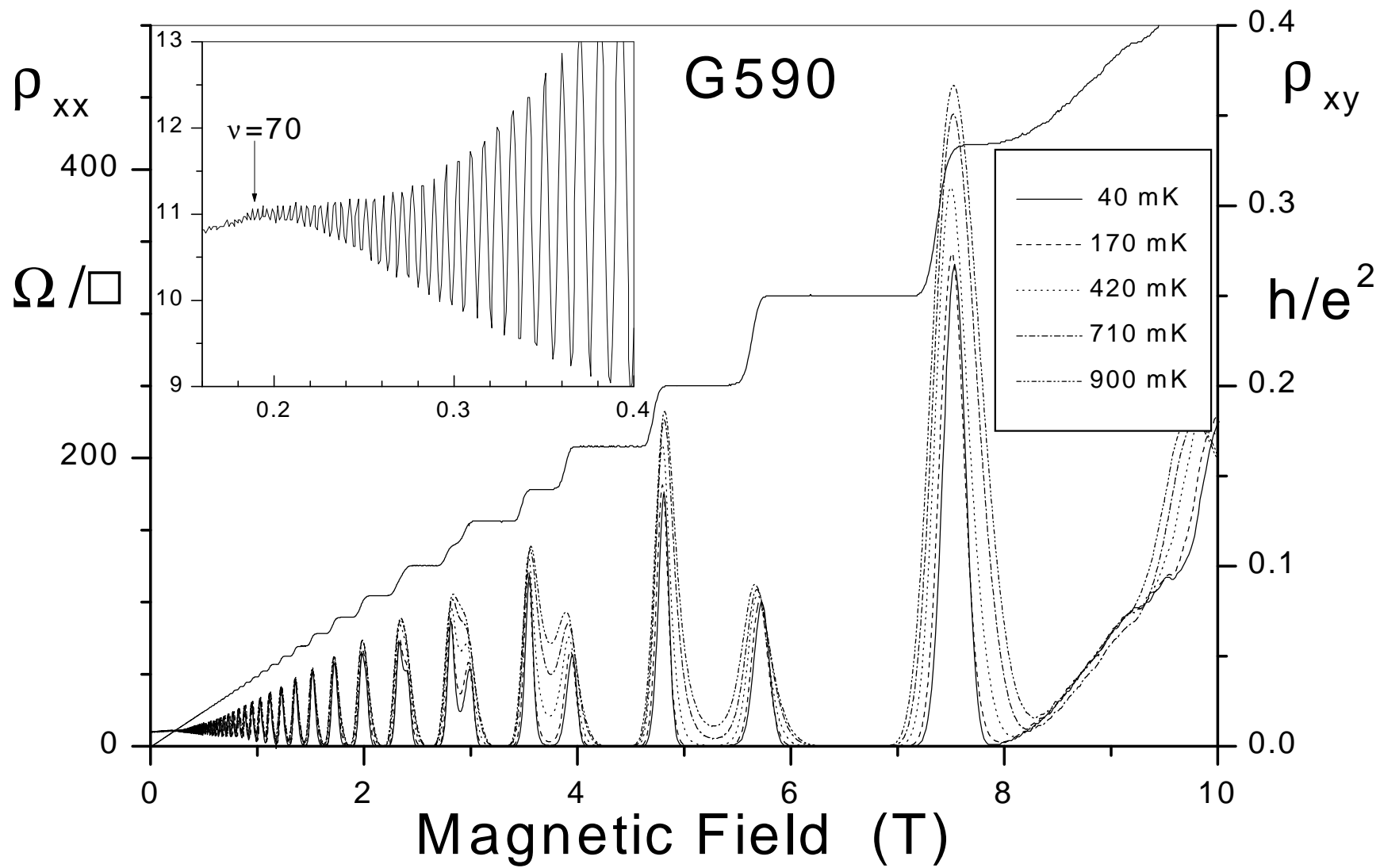
\begin{figure}
\caption{Magnetoresistance of sample G590 at several temperatures, showing a critical collapse 
of the spin splitting at $\nu=11$ (2.5~T).  The insert shows well resolved SdHOs to $\nu>70$.}
\label{fig:sdh590}
\end{figure}

The spin splitting disappears in quite a different way.  If the minima at odd $\nu$ just became 
unresolved at a certain field, it could be argued that the signal was becoming lost in the noise 
due to an exponential term, with a damping factor different from that of the regular SdHOs.  In 
this case spin splitting would be seen at lower fields (higher $\nu$) if the experiment were 
performed more carefully.  However, this does not appear to be the case and, in addition, there is 
a finite field where the maxima from either side of a spin split minimum converge, irrespective 
of the amount of experimental noise or resolution.  This is a sign that the collapse is critical, 
indicative of a second order phase change when the exchange interaction is turned off.  The 
position of the phase change is however dependent both on the sample and on the temperature.  
A dramatic example of the critical collapse may be seen for the higher density and more 
disordered sample G590 in Fig.~\ref{fig:sdh590}.  Here $\nu_c=11$ which is at a much higher 
magnetic field than for G641 (14 times greater), and shows only a very weak temperature 
dependence, changing from $\nu_c=11$ at 40~mK to $\nu_c=9$ at 900~mK.  This is however 
still a high quality sample as can be seen in the insert where the SdHOs are observed down to 
very low fields and only disappear into the noise below 0.2~T at around $\nu=70$.

The collapse of the spin splitting can be studied in at least three different ways: as a function of 
magnetic field at fixed temperature (preferably 0~K), as discussed above; for a given SdH peak 
(of Landau level $N$ \cite{N}) as function of temperature at fixed field, as will be discussed 
below; or as a function of density for a given SdH peak at fixed temperature.  We will only 
consider the first two cases.  

Wong {\em et al.}\ \cite{wong} used gated samples to study the density dependence and found 
that for each $N$ there was a critical density below which the splitting disappeared consistent 
with a phase change.  In their data, the density where $\delta\nu_N=0.5$ increased with the 
temperature of the measurement.  By fitting their data in the range $0.5<\delta\nu_N<0.9$ and 
extrapolating to $\delta\nu_N=0$ a critical density $n_c$ was found for each $N$ which 
allowed data from all peaks to be scaled onto a single curve for each sample.  Their results show 
that to first order $n_c\propto N$ which means that there was a critical magnetic field at which 
the spin splitting collapsed and the effect of varying the density was to align different filling 
factors with this critical field.  This is exactly what would be expected if disorder destroys the 
exchange enhancement whose energy scale is set by the magnetic field.  The critical field also 
varied between the samples, presumably in line with the disorder.  However, it would appear 
that changing density had little effect on the disorder, as otherwise the critical field would 
change with density and there would not be linear relationship between $N$ and $n_c$.  Hence, 
the vertical axis of the phase diagram depicted in Fig.~4 of Ref.~\cite{wong} is actually a 
measure of the spin gap size and not the disorder.

\section{Temperature dependence of spin splitting}

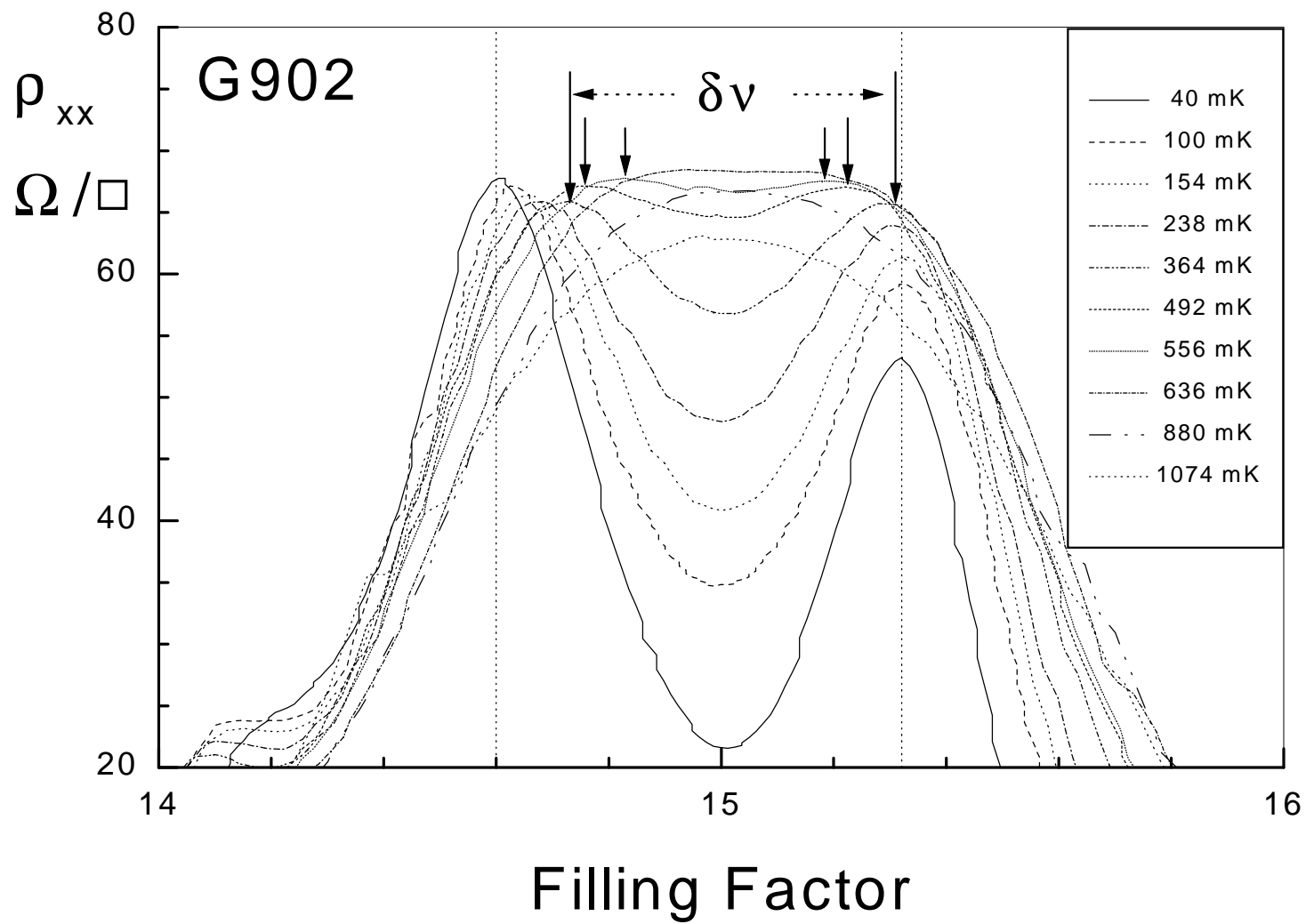
\begin{figure}
\caption{Spin splitting at $\nu=15$ for sample G902, showing how the maxima converge and 
the gap collapses as the temperature increases from 40~mK to 1~K }
\label{fig:dnu902}
\end{figure}

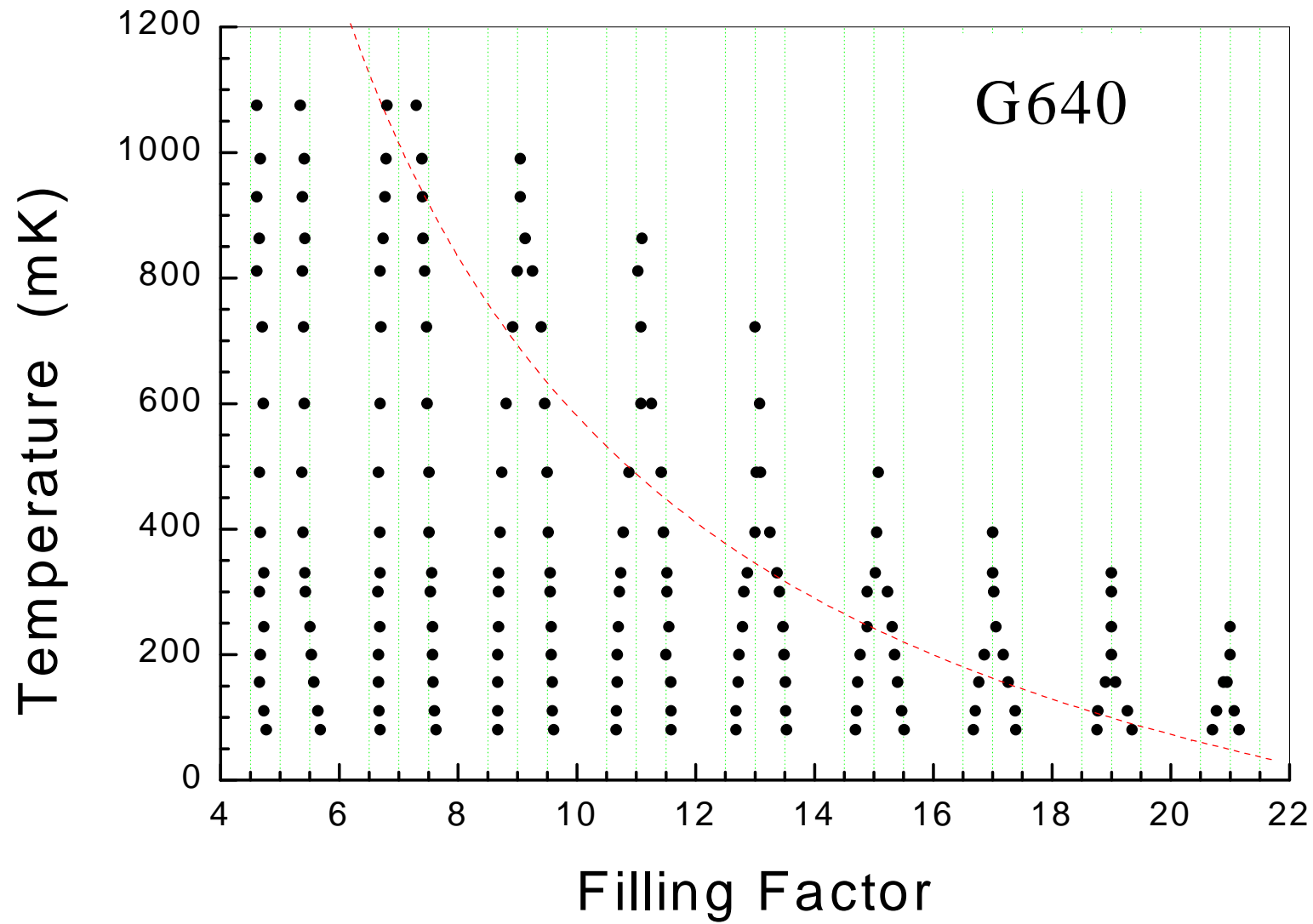
\begin{figure}
\caption{Temperature dependence of the filling factor at which spin split maxima occur in 
sample G640.  The dashed line, drawn through the positions where $\delta\nu=0.5$, defines the 
boundary between spin resolved and unresolved phases.}
\label{fig:maxima640}
\end{figure}

We will now consider how the spin splitting at a particular filling factor can be observed to 
collapse as a function of temperature.  Again it appears to be a critical phenomenon.  The 
temperature evolution of the resistivity around $\nu=15$ for sample G902 is shown in 
Fig.~\ref{fig:dnu902}. (This example is chosen as it shows the fully resolved spin split 
minimum at low temperature changing to a maximum by 1~K.)  At the lowest temperatures the 
two maxima can be seen at $\nu=15\pm0.4$ and as the temperature is increased these remain at 
the same filling factor while the minimum becomes shallower. Only when the depth of the 
minimum is $<10$\% of the peak height do the maxima start to converge and then they do so 
rapidly as $\delta\nu$ collapses. It will also be noticed that once less than 0.5, $\delta\nu$ 
becomes difficult to measure reliably as the peak has a rather flat top. 
Figure~\ref{fig:maxima640} illustrates the actual filling factors at which maxima occur for each 
temperature (this time for sample G640 but all the samples behave similarly). The dotted lines 
on this figure show the positions expected when the spin splitting is either completely resolved 
or completely absent.  By their convergence, the points clearly show transitions between these 
two limiting cases at different temperatures for each filling factor. Figure~\ref{fig:maxima640} 
thus represents the temperature driven phase diagram for sample G640 with the spin resolved 
phase to the lower left side of the figure.  The dashed line on the figure, drawn through the 
positions where $\delta\nu=0.5$, defines the phase boundary.  

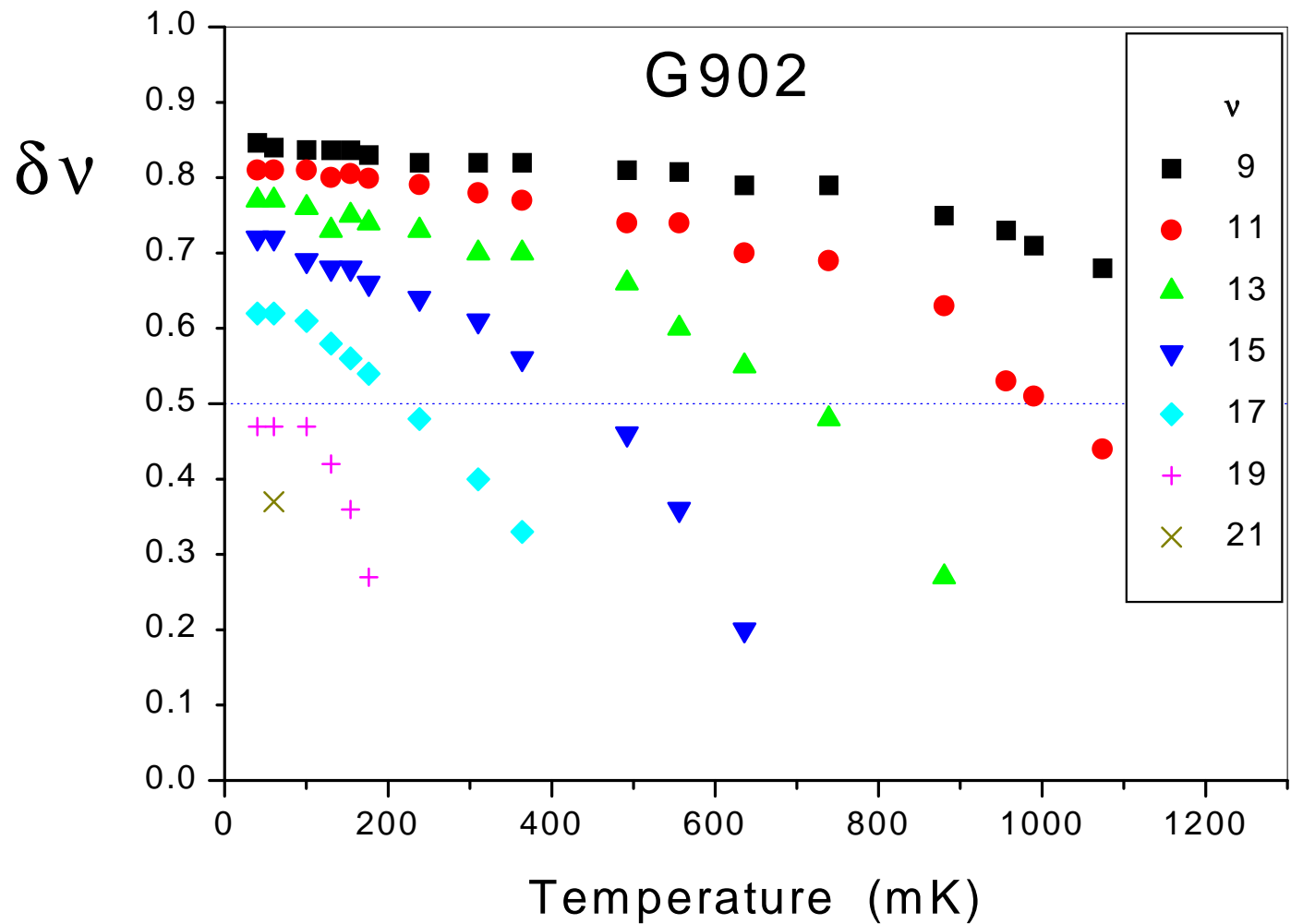
\begin{figure}
\caption{Separation of each spin split maxima $\delta\nu$ for sample G902 as a function of 
temperature.}
\label{fig:dnuT902}
\end{figure}

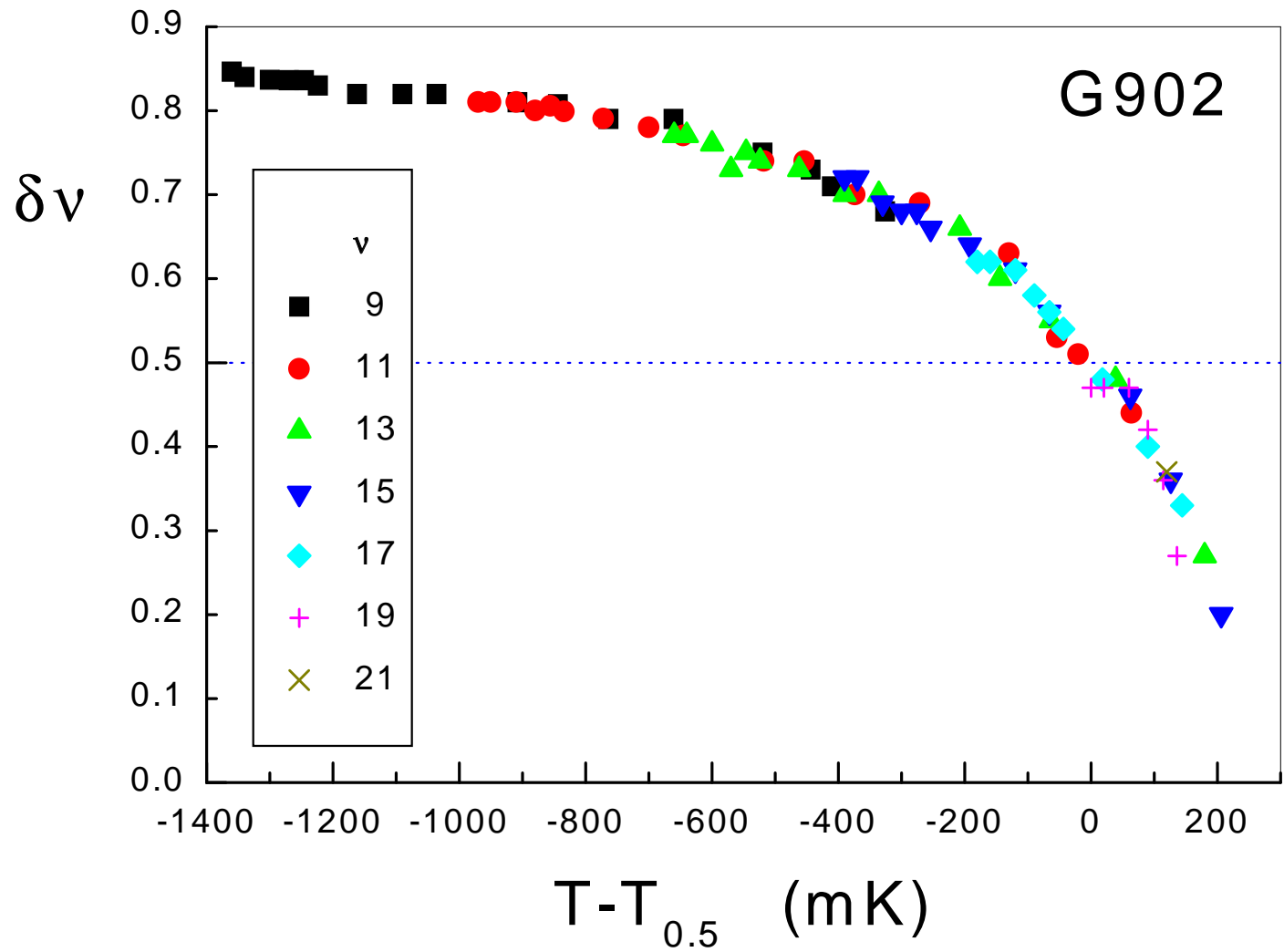
\begin{figure}
\caption{The data from the previous figure showing a single curve as a function of $T-
T_{0.5}$, where $T_{0.5}$ is the temperature at which $\delta\nu=0.5$ at each filling factor.}
\label{fig:scaled902}
\end{figure}

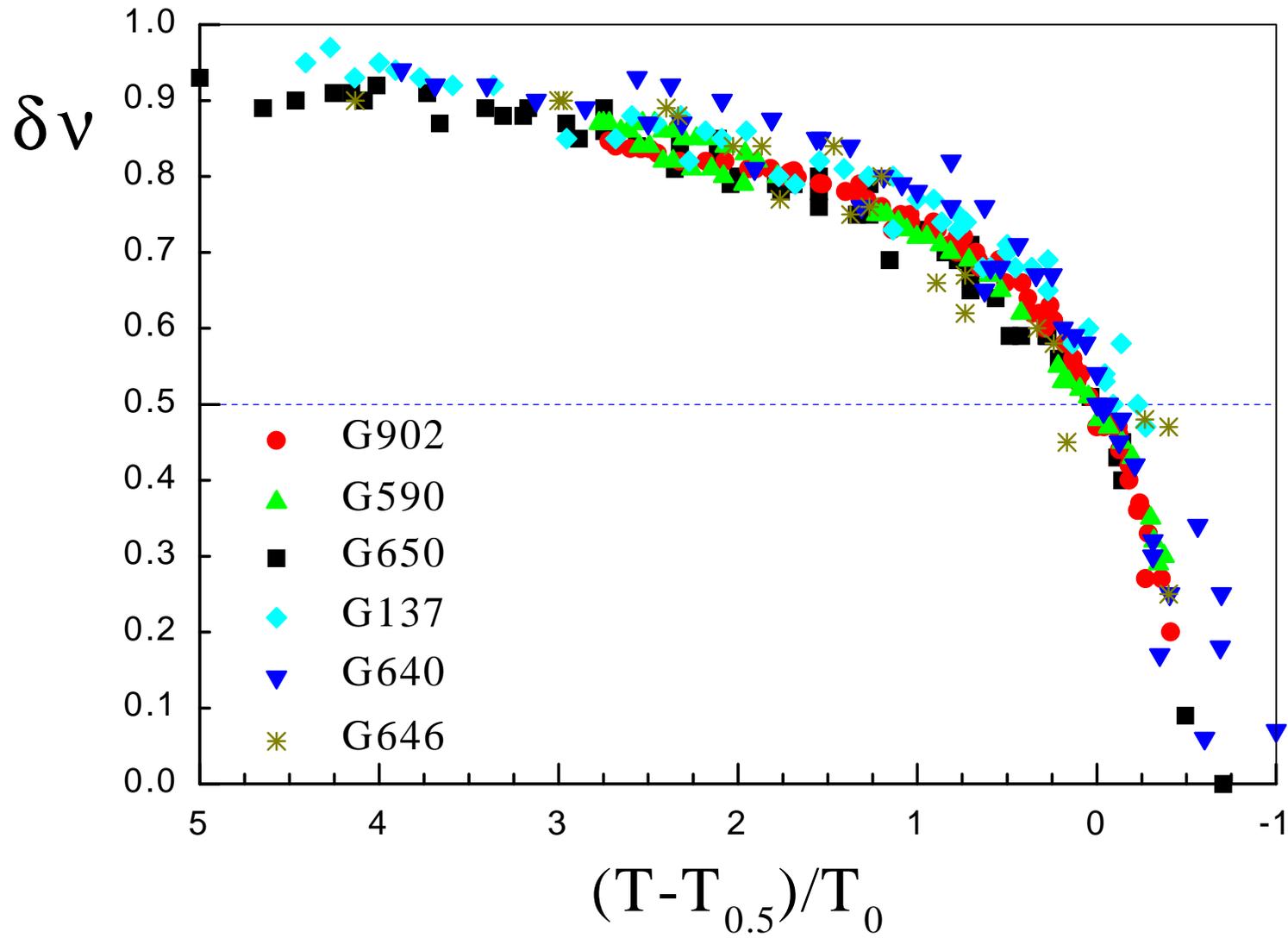
\begin{figure}
\caption{Universal behavior of $\delta\nu$ for all samples and all $\nu$ appears with the 
temperature suitably scaled. $T_0$ is a sample dependent scaling parameter, which corresponds 
to the temperature range over which the spin gap collapses.}
\label{fig:dnuscaled}
\end{figure}

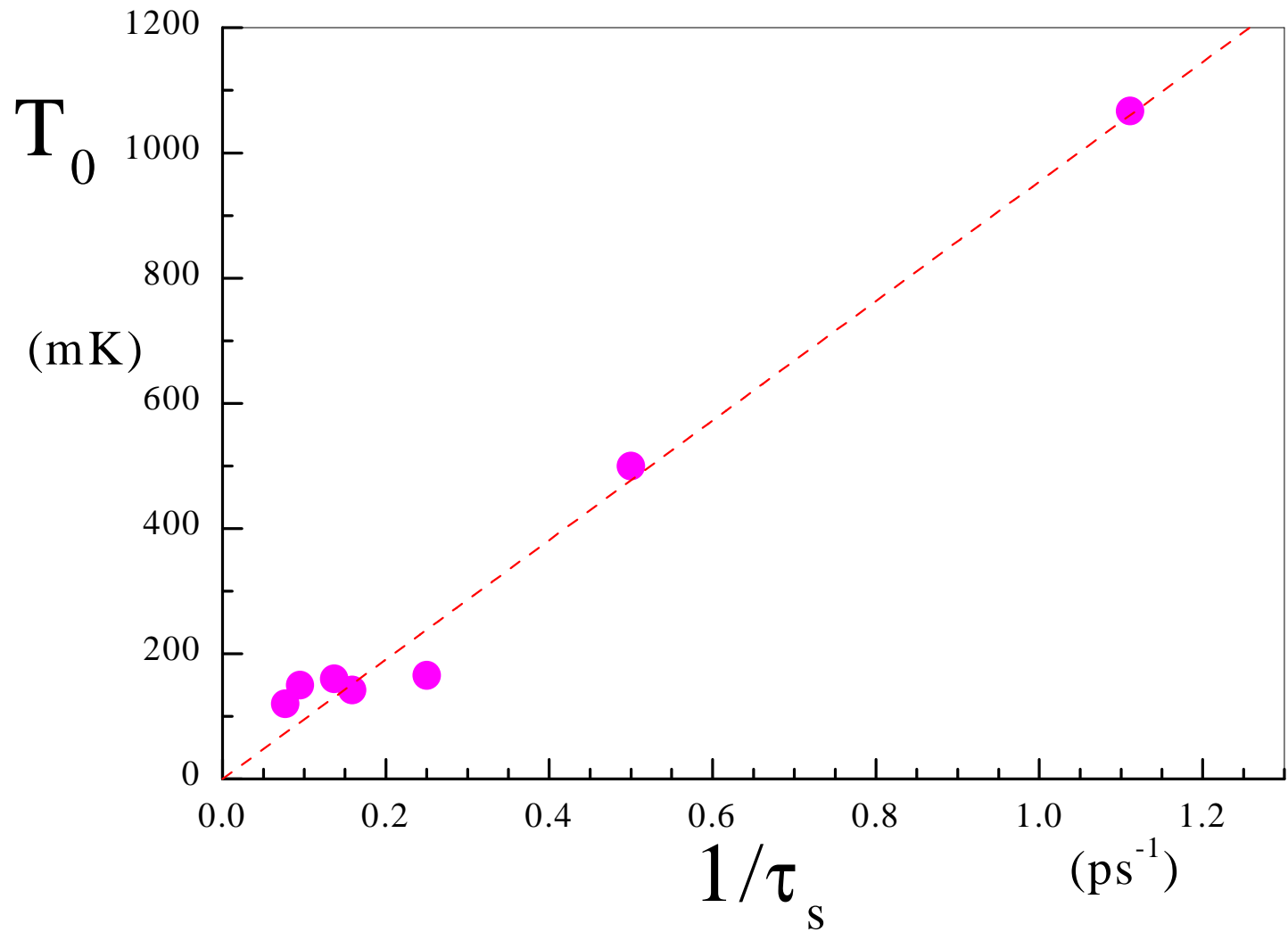
\begin{figure}
\caption{Variation of the sample dependent scaling temperature $T_0$ with disorder, 
parameterised as the inverse quantum lifetime.}
\label{fig:T0}
\end{figure}

The separation of the individual spin split maxima is displayed as a function of temperature in 
Fig.~\ref{fig:dnuT902} for odd filling factors in the range $19<\nu<9$ from sample G902.  For 
each filling factor the collapse looks to be quite similar but occurs at a different temperature.  At 
lower filling factors there is no noticeable change in $\delta\nu$ in the temperature range of the 
experiment, although there would again be a collapse at higher temperature.  If the spin gap had 
just one component that increased with magnetic field then it should be possible to collapse the 
data of Fig.~\ref{fig:dnuT902} onto a single curve as a function of a scaled temperature i.e.\ 
$\delta\nu(T/T_{\nu})$ with a $T_{\nu}$ determined for each filling factor.  However, the data 
does not scale in this way.  This can be seen by looking at the maximum value of $\delta\nu$, 
which would be 1.0 at $T=0$ for all $\nu$ if $\delta\nu$ were a function of $T/T_{\nu}$, but in 
fact decreases at higher $\nu$.  Instead, the data can be collapsed by simply shifting the 
temperature axes, i.e.\ plotting it as $\delta\nu(T')=\delta\nu(T-T_{0.5})$ in 
Fig.~\ref{fig:scaled902}, where $T_{0.5}$ is the temperature where $\delta\nu=0.5$ for each 
filling factor. $T_{0.5}$ is used in preference to the temperature where $\delta\nu=0$ as it is 
both experimentally accessible and corresponds to the mid-point of the gap's collapse.  No 
fitting of the data is required so we do not have to assume any functional form to the collapse of 
the gap, we just read off $T_{0.5}$ at each filling factor from Fig.~\ref{fig:dnuT902}.  The 
remarkable behavior shown in Fig.~\ref{fig:scaled902} demonstrates that the gap collapses over 
the same range of temperature for all filling factors.  In other words there is a single finite width 
to the phase boundary.  This suggests that once there is sufficient thermal energy to initiate a 
collapse (which increases with magnetic field) the same amount of additional thermal energy 
will complete the phase transition at all magnetic fields.  Other samples show a similar behavior 
but with the transition occurring over a smaller temperature range in higher mobility samples, 
i.e.\ the width of the phase boundary decreases as the mobility increases.  Indeed it is possible to 
collapse the data for all the samples (and all filling factors) onto a single curve by normalizing 
$T-T_{0.5}$ by a sample dependent temperature $T_0$ as shown in Fig.~\ref{fig:dnuscaled}.  
The value used for $T_0$ is found by fitting the data for each sample to 
$\delta\nu=0.5+a_1\bigl(1-\exp\left((T-T_{0.5})/T_0\right)\bigr)$, which is the simplest 
function that provides a reasonable fit the data.  Similar values of $T_0$, but slightly lower 
quality fits, can be obtained from a modified Brillouin function, as used in Ref.~\cite{wong}, 
$\delta\nu=a\coth\left(2a(T-T_{0.5}-T_0)/T_0\right)-b\coth\left(2b(T-T_{0.5}-
T_0)/T_0\right)$.  At present we do not understand the physical significance of either of these 
functional forms, merely using them to extract $T_0$, but we do note that $T_0$ is a good 
measure of the temperature range over which the transition proceeds.  Figure~\ref{fig:T0} 
shows how $T_0$ varies with disorder in the samples, where the degree of disorder is 
represented by the inverse quantum lifetime $1/\tau_s$, measured from the low field SdHOs in 
each sample.  This assumes that the SdHO broadening is due to an impurity potential 
$\sim\hbar/\tau_s$.  Although there is a lot of scatter on the graph (due to problems in 
measuring $\tau_s$ and the long route to finding $T_0$) it shows a clear correlation between 
$T_0$ and disorder.  The linear fit shown has a gradient of 0.95~K~ps, which means 
$\hbar/\tau_s=8.0 k_BT_0$.  We note the similarity between this factor and the fact that 
Eq.~(\ref{eqn:LK}) predicts SdHO minima will be $\sim50\%$ developed when $k_BT\sim 
E_g/6$.  This appears to confirm a connection between the collapse of the spin splitting and the 
disorder potential.

\section{Energy gaps at odd integer filling factors}

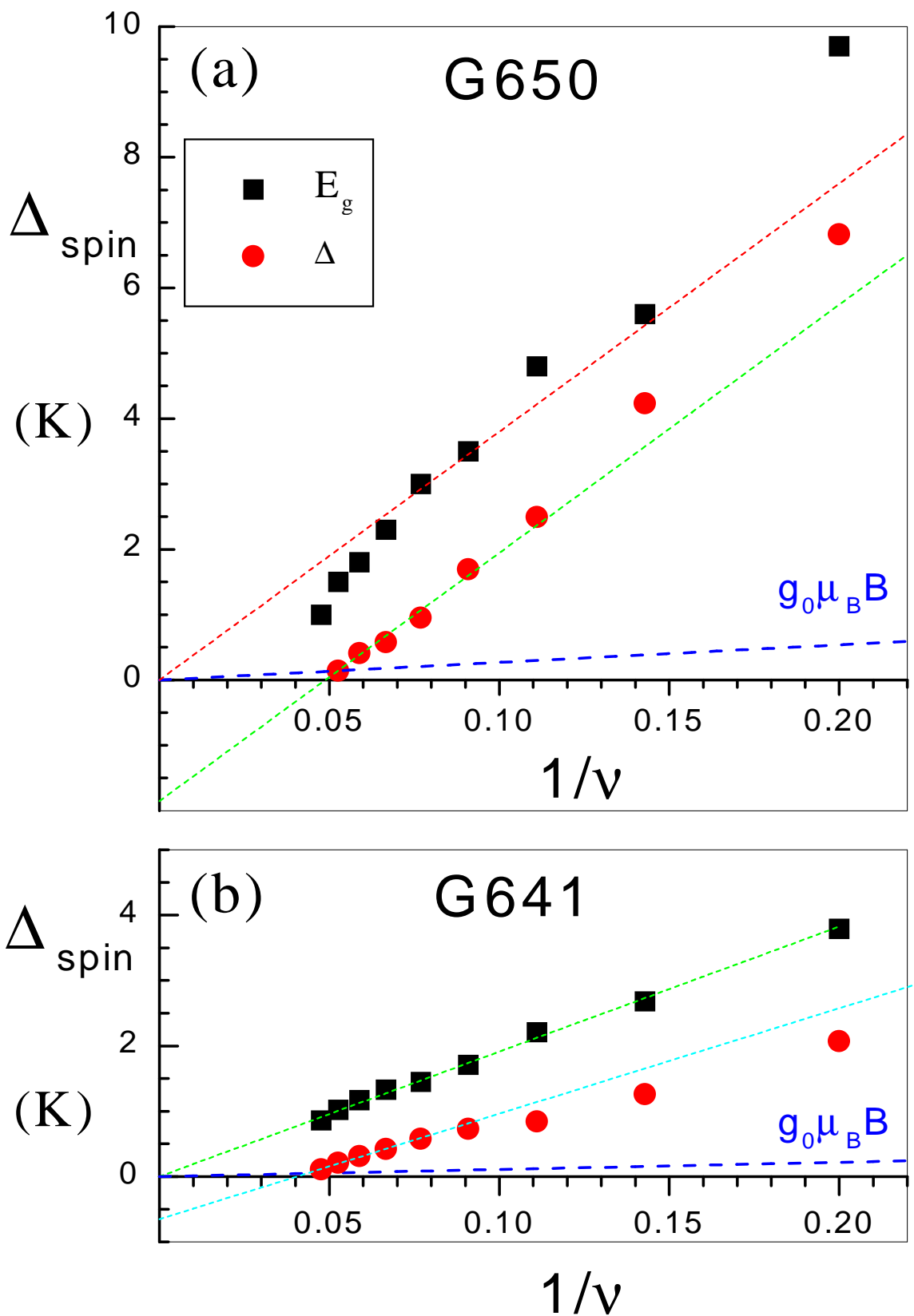
\begin{figure}
\caption{Energy gaps at odd integer filling factor for samples (a) G650 and (b) G641.  The fitted 
lines are discussed in the text. The dashed line is the single particle Zeeman energy.}
\label{fig:Enu}
\end{figure}

We now turn our attention from the spin-split maxima to the minima at odd integral $\nu$ and 
use the temperature dependence of the resistivity there to evaluate the energy gaps at each odd 
filling factor.  As we have previously discussed the data can be analyzed in two ways. Either the 
actual value of $\rho_{xx}$ at the minimum can be used and an activation energy $\Delta$ 
extracted from an Arrhenius plot, or the depth of the minimum can be used to obtain an energy 
gap $E_g$ by fitting to the Lifshitz-Kosevich formula Eq.~(\ref{eqn:LK}). $\Delta$ and $E_g$ 
represent the energy gaps between mobility edges and between Landau level centers respectively 
and we expect that $E_g=\Delta +\Gamma$, where $\Gamma$ is the width of the region of 
extended states.  Both $\Delta$ and $E_g$  are shown on Fig.~\ref{fig:Enu} for samples G641 
and G650.  The dashed line on this figure is the single particle Zeeman energy $g_0\mu_BB$.  
That this is much smaller than the measured energy gaps shows the gaps are dominated by 
exchange energy.  Although there are quite large experimental uncertainties in the measured 
gaps, especially at large $\nu$, it can be seen that $\Delta$ is approximately linear in $1/\nu$ 
(i.e. magnetic field) and becomes zero at a finite filling factor, giving us a measure of $\nu_c$ 
where the spin gap closes.  (These values of $\nu_c$ will be discussed in the next section.)  The 
values of $E_g$ measured at magnetic fields above this critical region also appear to be linear in 
$B$, with the same slope as $\Delta$ but this time extrapolating to the origin.  This is consistent 
with a constant value of $\Gamma$ that can be obtained from the negative intercept of the fit to 
$\Delta$.  Near the critical region there is some indication that $E_g$ starts to decrease below 
the straight line but unfortunately it becomes unmeasureable once $\delta\nu<0.5$ where there is 
no longer a clear minimum observable over a range of temperatures.  Furthermore, when the gap 
is collapsing as a function of temperature (as discussed above) the temperature dependence of 
resistivity can not be used to obtain a value for the gap.  We are thus limited to data at 
temperatures below the point where the maxima start to converge.  For $\nu<<\nu_c$ this is not 
a problem, but it increases the uncertainty in $E_g$ when $\nu\sim\nu_c$ as there are then 
fewer suitable points for fitting.

It is important to note that these measurements suggest an energy gap due to exchange that 
increases {\em linearly} with magnetic field and is {\em independent} of temperature once the 
enhancement is turned on.  This allows us to use an effective $g$-factor that is the same at all 
fields, and to describe the spin gap as $\Delta_{spin}=g^*\mu_BB$.  Furthermore $g^*$ can be 
obtained from the gradients of the fits shown in Fig.~\ref{fig:Enu} as 6.2 for G650 and 7.6 for 
G641, values that are generally in agreement with earlier work \cite{usher}. Many workers have 
previously reported such a linear increase \cite{usher,dolgo} and found their results puzzling, 
since simple theories of exchange suggest the gap should increases like $\sqrt{B}$.  This square 
root dependence arises when the magnetic length $l_B=\sqrt{\hbar/eB}$ represents the average 
separation of electrons which determines the Coulomb energy 
$E_c=e^2/4\pi\epsilon_0\epsilon_r l_B$.  However, this is only strictly true when there is just 
one Landau level occupied and the Coulomb energy is small compared with the cyclotron 
energy.  When there are many LLs occupied neither of these conditions are fulfilled: 
$E_c>\hbar\omega_c$ and only $1/\nu$ of the electrons are able to contribute to the exchange 
energy as the others are in full LLs.  At zero field, and also in the limit of large $\nu$, the 
correlation length will be set by the Fermi wavevector $k_F$ rather than $L_B$.  These two 
factors lead to an exchange energy $E_{ex}\sim \left(e^2/4\pi\epsilon_0\epsilon_r\right) 
\left(k_F/\nu\right)$ which depends on density and increases linearly with the magnetic field at 
which each filling factor falls.  Thus we can write 
\begin{equation}
E_{ex}=\alpha\hbar\omega_c
\end{equation}
and since $\mu_B=e\hbar/2m_0$, with $m_0$ the free electron mass and $m^*$ the effective 
mass, $\alpha$ is related to the $g$-factor by:
\begin{equation}
\alpha=(g^*-g_0)\frac{m^*}{2m_0}
\end{equation}
Aleiner and Glazman \cite{aleiner} have used a more sophisticated Hartree-Fock calculation 
with Thomas-Fermi screening to show that exchange energy should increase linearly with 
magnetic field in the large filling factor limit, but that there is an additional logarithmic factor 
such that $\alpha=\left(1/\pi k_F a_B\right) \ln(2k_F a_B)$, where $a_B$ is the effective Bohr 
radius.  Thus $\alpha$ is predicted to vary with carrier density since, with $n_e$ in units of 
$10^{15}m^{-2}$, $k_F a_B=1.13\sqrt{n_e}$ giving a value of $\alpha=0.23$ at 
$n_e=10^{15}m^{-2}$.  While FS use this value of $\alpha$ in their calculation they do not 
seem to account for its density dependence.

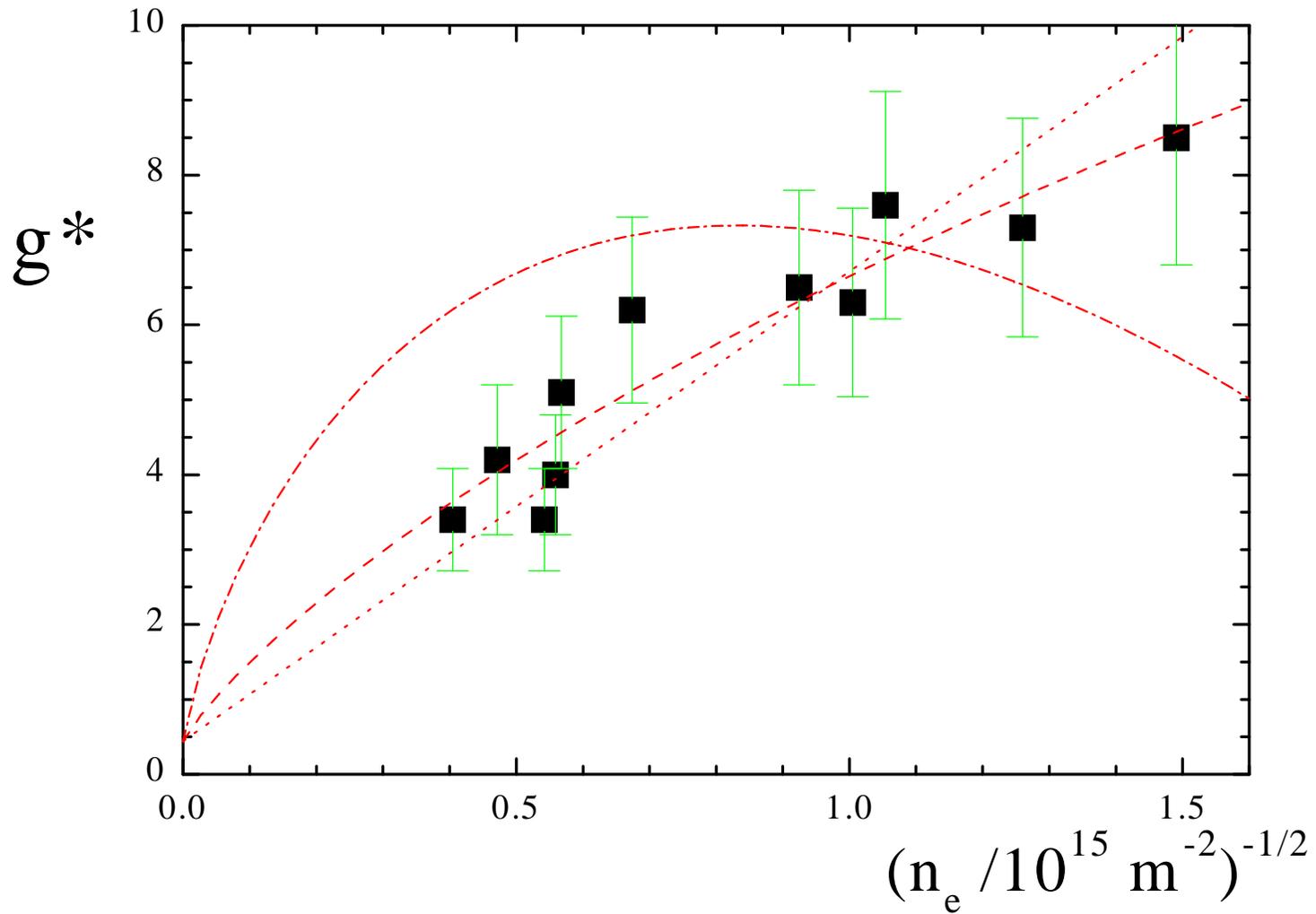
\begin{figure}
\caption{Density dependence of the measured effective $g$-factor.  The dotted and dashed 
curves are fits to the data described in the text and the dot-dashed line is the theory of Ref. 15}
\label{fig:gvsn}
\end{figure}

The values of $g^*$ obtained for each sample from the gradients of plots such as 
Fig.~\ref{fig:Enu} are shown as a function of $n_e^{-1/2}$ in Fig.~\ref{fig:gvsn}.  This figure 
clearly shows that there is a density dependence of $g^*$ and that it is somewhat weaker than 
$n_e^{-1/2}$.  The dotted and the dashed lines given by the equations 
$g^*=g_0+6.27/\sqrt{n_e}$ and $g^*=g_0+6.22/\sqrt{n_e}+1.88 \ln(\sqrt{n_e})/\sqrt{n_e}$ 
represent the simplest and the best fits to the data respectively, which have been constrained to 
pass through $g^*=g_0$ at large density.  By contrast the dash-dotted line is the prediction of 
Ref. \cite{aleiner} ($g^*=g_0+6.75/\sqrt{n_e}+8.29\ln(\sqrt{n_e})/\sqrt{n_e}$), which does 
not fit the data particularly well and produces an unexpected maximum.  The discrepancy is 
mostly due to an over estimate of the logarithmic correction, and apart from this numerical 
factor the experiment and theory are in reasonable agreement.

\section{The critical filling factor --- $\nu_c$}

Having discussed something of the temperature dependence of the spin splitting we can now 
establish a reliable way of obtaining the critical filling factor $\nu_c$ at $T=0$, which is the 
quantity required for comparison with the theory of FS.  The first, and most direct, approach is 
to follow the temperature dependence of the separation $\delta\nu$ between the spin split 
maxima.  The crudest method of finding $\nu_c$ is by extrapolation of the temperatures where 
$\delta\nu=0.5$ on Fig.~\ref{fig:maxima640} to $T=0$.  The problem with this approach is that 
the value obtained is sensitive to the functional form chosen for the curve, particularly if the 
experimental data does not extend to sufficiently low temperature.  

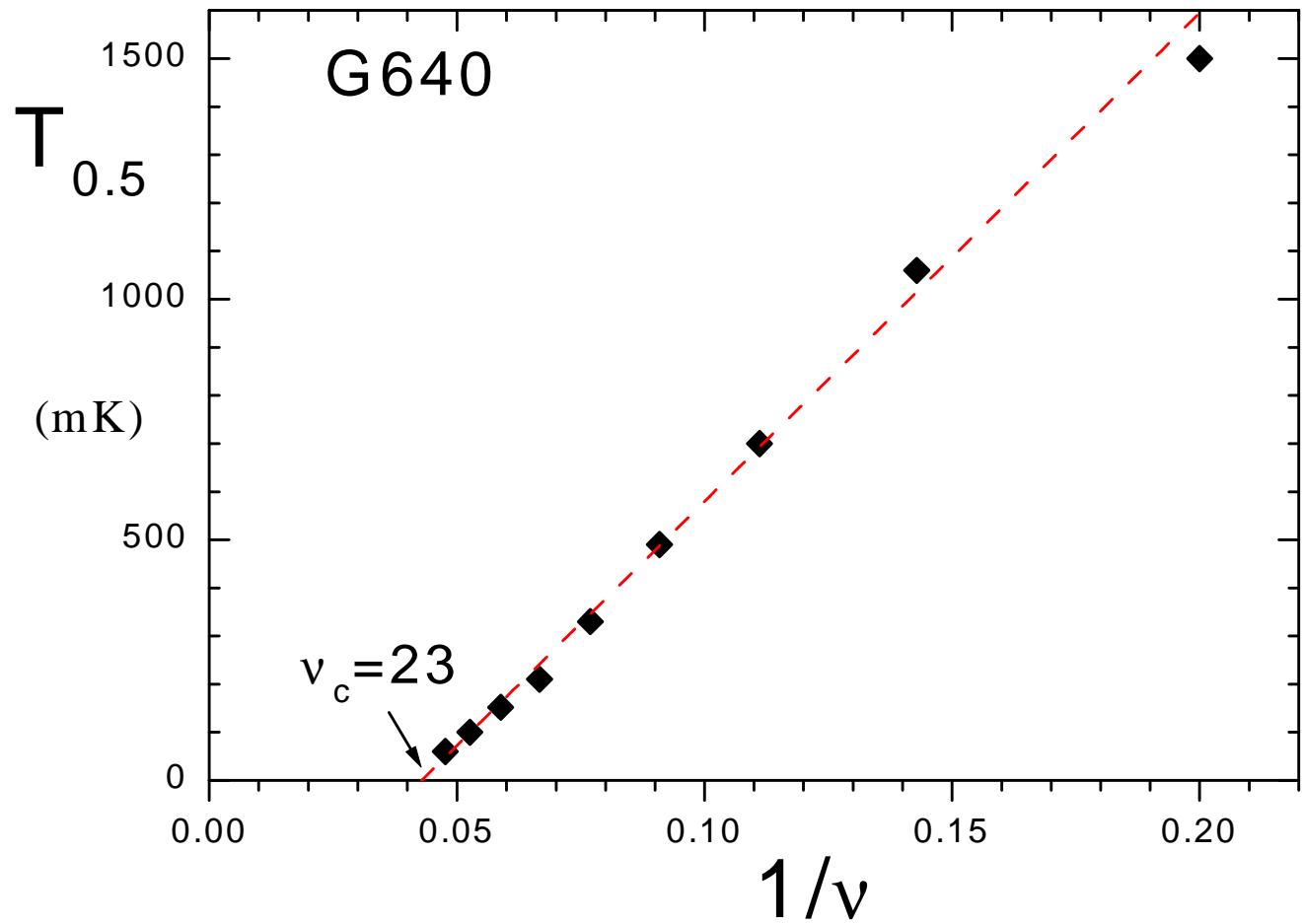
\begin{figure}
\caption{Variation of $\nu_c$ with temperature for sample G640.  The intercept of the straight 
line fit gives the critical filling factor at $T=0$.}
\label{fig:nuc640}
\end{figure}

A more accurate method is to plot the values of $T_{0.5}$ as a function of $1/\nu$ and 
extrapolate to $T=0$ as shown in Fig.~\ref{fig:nuc640}.  This graph shows a linear dependence 
with the intercept giving the critical filling factor at $T=0$, of 23 for this sample.  From these 
data it is quite clear that at $T=0$ the spin splitting will collapse at a certain magnetic field as 
opposed to tailing off exponentially.  A similar approximately linear relationship is observed for 
all the samples and so we propose the empirical relationship: 
\begin{equation}
\frac{1}{\nu_c(T)}=\frac{1}{\nu_c(0)}+cT
\label{eqn:nucT}
\end{equation}
with $c$ and $\nu_c(0)$ as sample dependent parameters.  These parameters will be discussed 
in more detail later in the paper, but at this point it is important to note that there is no direct 
correlation between them.  This means that $\nu_c(0)$ can not be obtained just by finding the 
last spin split maximum in a single finite temperature resistivity trace.  There is sometimes a 
deviation from linearity at the lowest filling factors $\nu=3$ or 5, with values of $T_{0.5}$ up 
to 20\% below the line.  This may be an artifact of the experiment since the transition region 
occurs above 1~K where the temperature in the fridge was not always stable.  Alternatively it 
may provide evidence for skyrmionic excitations at these filling factors having a 
correspondingly lower energy than the single spin flips.  In any case we have not used these 
points in our analysis of $\nu_c$.

The critical filling factor can also be obtained from the point where the mobility gap $\Delta$ 
collapses, via the temperature dependent resistivity data discussed in the previous section.  The 
values of $\nu_c$ obtained from the two methods are very similar, although for the cleanest 
samples the collapsing of the gap gives slightly smaller values.  This may be due to the difficulty 
in measuring small energy gaps and for the remainder of the paper we will use the former 
method.

\subsection{Sample dependence of $\nu_c$}

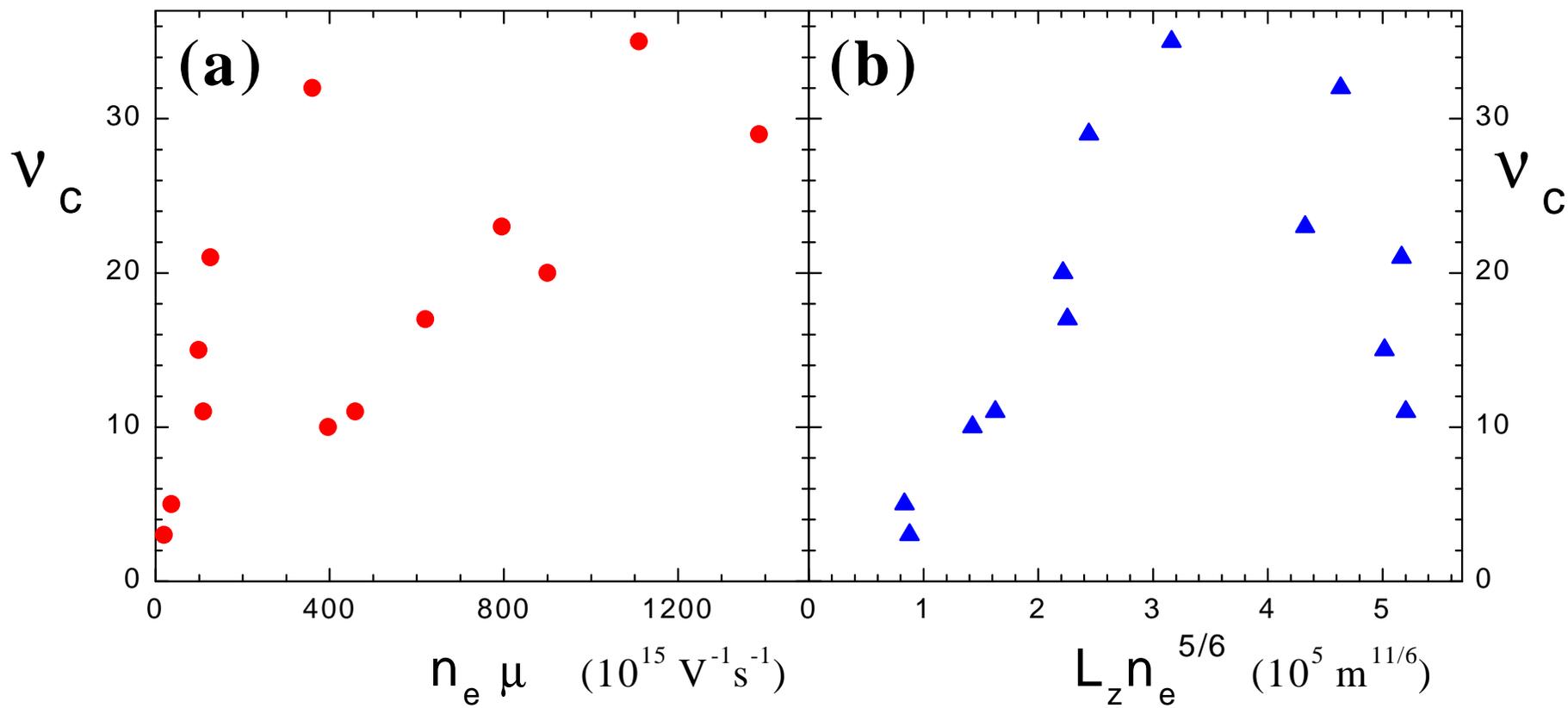
\begin{figure}
\caption{Test of the dependence of $\nu_c$ on the sample parameters predicted by FS and 
discussed further in the text.}
\label{fig:nuln}
\end{figure}

Having established a method of finding the $T=0$ value of $\nu_c$ we can now investigate how 
it varies between samples.  In their calculations Fogler and Shklovskii \cite{fs} found different 
expressions for the dependence of $\nu_c$ on sample parameters according to the range of the 
dominant scattering mechanism in each case.  For low mobility samples they found $N_c\propto 
n_e\mu$ and for high mobility samples $N_c\propto L_z n_e^{5/6}/n_i^{1/3}$, with electron 
densities in our range of interest and similar forms of expressions for other densities.  
These predictions are tested by plotting our measured $\nu_c$ against $n_e\mu$ and $L_z 
n_e^{5/6}$ in Fig.~\ref{fig:nuln}.  Clearly there are some samples that agree with the 
predictions in each case but neither description applies to all the samples and there is no clear 
distinction between the low and high mobility samples.  This lack of agreement is not 
unexpected since our samples cover a wide range of densities and mobilities and different 
scattering mechanisms will dominate.  In drawing these figures we have also to assume that the 
impurity density $n_i$ is constant, as there is no way of measuring this directly, but in practice 
it may differ widely between the samples.  However, in FS all the sample parameters enter 
through the calculation of the scattering rate $\tau_s$.  Rather than try to calculate this quantity 
we will show that the theory is essentially correct by using the measured value of $\tau_s$, 
obtained from the SdHOs at low magnetic field.

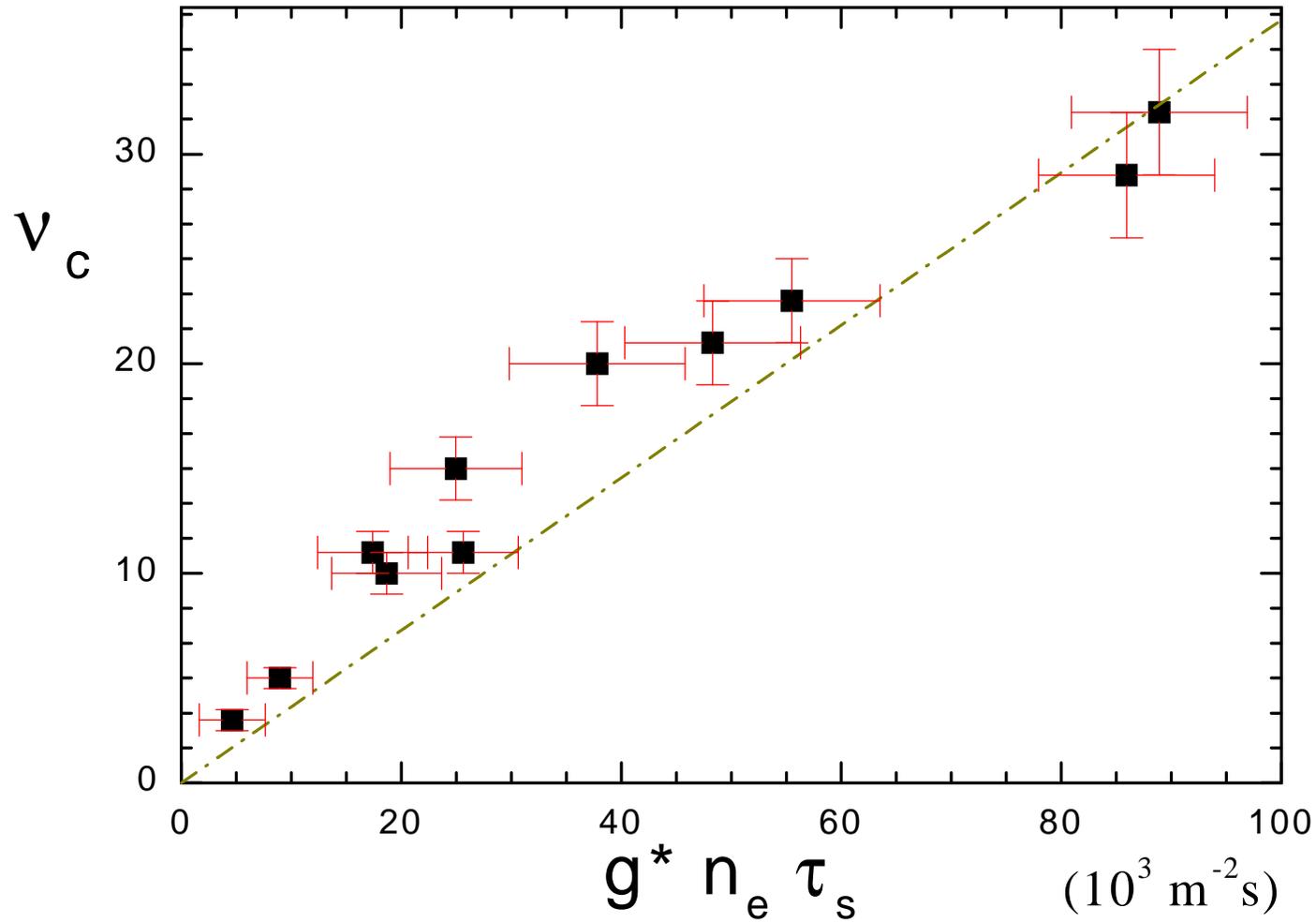
\begin{figure}
\caption{The critical filling factor as a function of the product $g^*n_e\tau_s$, obtained from 
measuring each parameter, showing excellent agreement with the prediction of Eq. (7).}
\label{fig:nugntau}
\end{figure}

Simply speaking the spin splitting will collapse when the energy separation of spin up and down 
levels is less than their disorder broadening {\em i.e.}\ when 
\begin{equation}
g^*\mu_BB=\hbar/\tau_s .
\end{equation}
At this point the exchange contribution will disappear and only the small bare Zeeman splitting 
will remain.  Rearranging this equation in terms of the critical filling factor and electron density 
we obtain
\begin{equation}
\nu_c=g^*n_e\tau_s\frac{h}{2m_0}
\label{eqn:nuc}
\end{equation}
The $T=0$ values of $\nu_c$ are plotted as a function of $g^*n_e\tau_s$ in 
Fig.~\ref{fig:nugntau}, together with the prediction of Eq.~(\ref{eqn:nuc}).  Clearly this 
provides a good description for all the samples without using any adjustable parameters.  There 
are quite large experimental uncertainties in the measurement of both $g^*$ and $\tau_s$ 
indicated by the error bars.  (A reliable value of $\tau_s$ could not be obtained for sample G627 
because the magnetic field was swept too fast to resolve the SdHOs at low field.  However, 
$\nu_c$ for this sample is quite in line with the value expected.)  It is quite remarkable how well 
this very simple approach matches the experimental data particularly that the disorder 
broadening seems to be adequately described by $\hbar/\tau_s$ without any numerical factors.

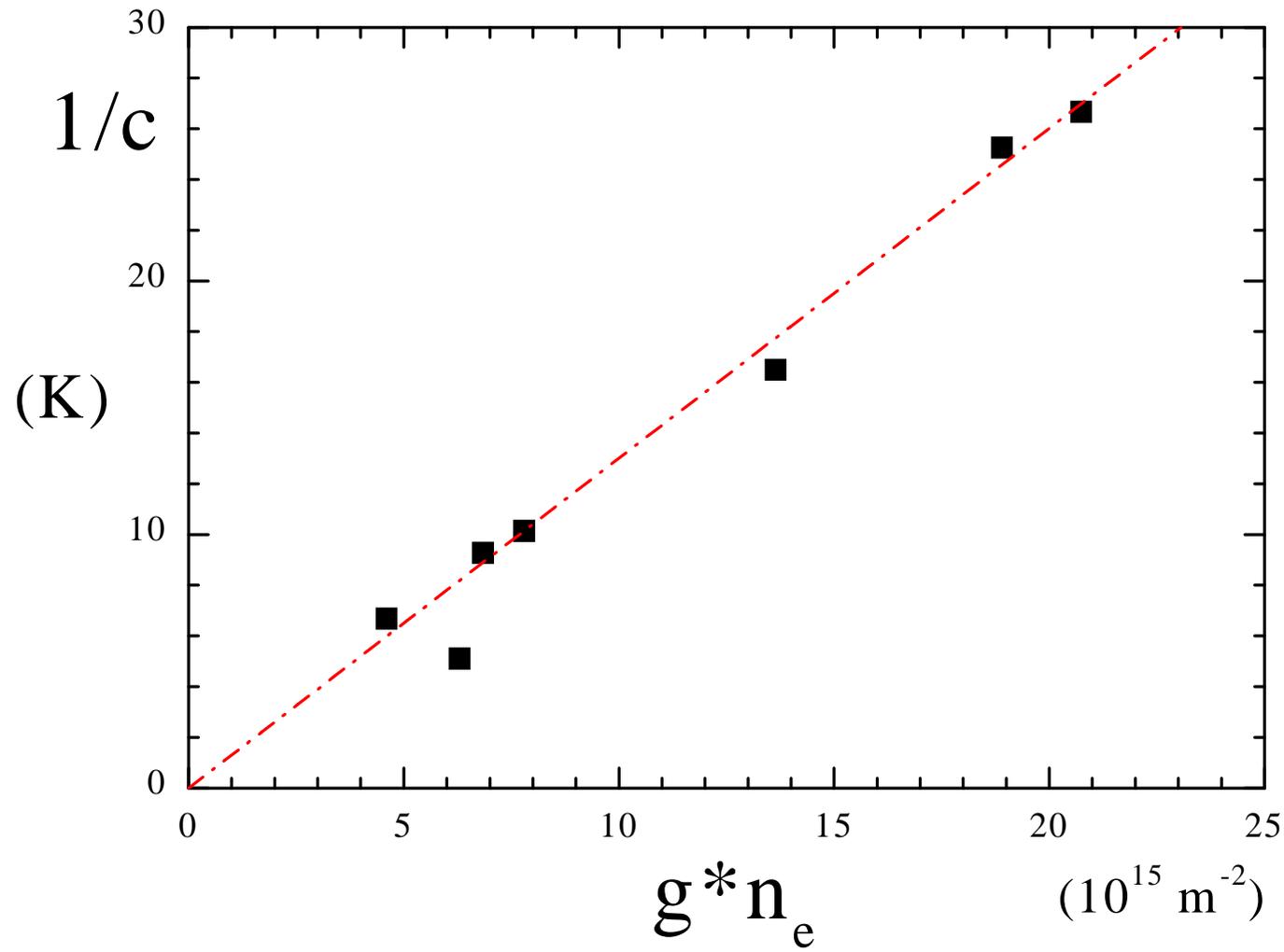
\begin{figure}
\caption{Sample dependance of the parameter $c$, determined from the gradients of plots like 
Fig. 12.}
\label{fig:cgn}
\end{figure}

Returning to Eq.~(\ref{eqn:nucT}), it can be seen that not only did the critical filling factor at 
$T=0$ vary between the samples, but so did the rate of change of critical filling factor with 
temperature.  We will now try to understand this temperature dependence in another simple 
model.  As the temperature is increased, reversed spins will be thermally excited, which will 
reduce the exchange correlation.  Thus the exchange enhancement can be destroyed even when 
$g^*\mu_BB>\hbar/\tau_s$.  We propose the critical condition for the spin splitting to collapse 
at finite temperature to be
\begin{equation}
g^*\mu_BB=\hbar/\tau_s + M k_BT
\end{equation}
where $M$ is a numerical constant that determines the effect of the thermal fluctuations on the 
exchange energy.  Comparing this with Eq.~(\ref{eqn:nucT}), and using Eq.~(\ref{eqn:nuc}) to 
replace $\tau_s$ with $\nu_c(0)$, shows that the parameters $c$ and $M$ that we have 
introduced are related by $c=M k_B m_0/\pi\hbar^2 n_e g^*$, {\em i.e.}\ just fundamental 
constants and the sample dependent product $n_eg^*$.  In Fig.~\ref{fig:cgn} the linear 
dependence of $1/c$ on this product is very clear and from the slope of this graph we obtain a 
value of $M=2.1$ which appears quite reasonable.  Again it is remarkable how well this simple 
model, of thermally excited spins aiding the disorder potential in destroying the exchange 
correlation, accounts for the data.  A full theory is required to account for the actual value of 
$M$.

\section{Tilted magnetic field and the enhanced \lowercase {g}-factor}

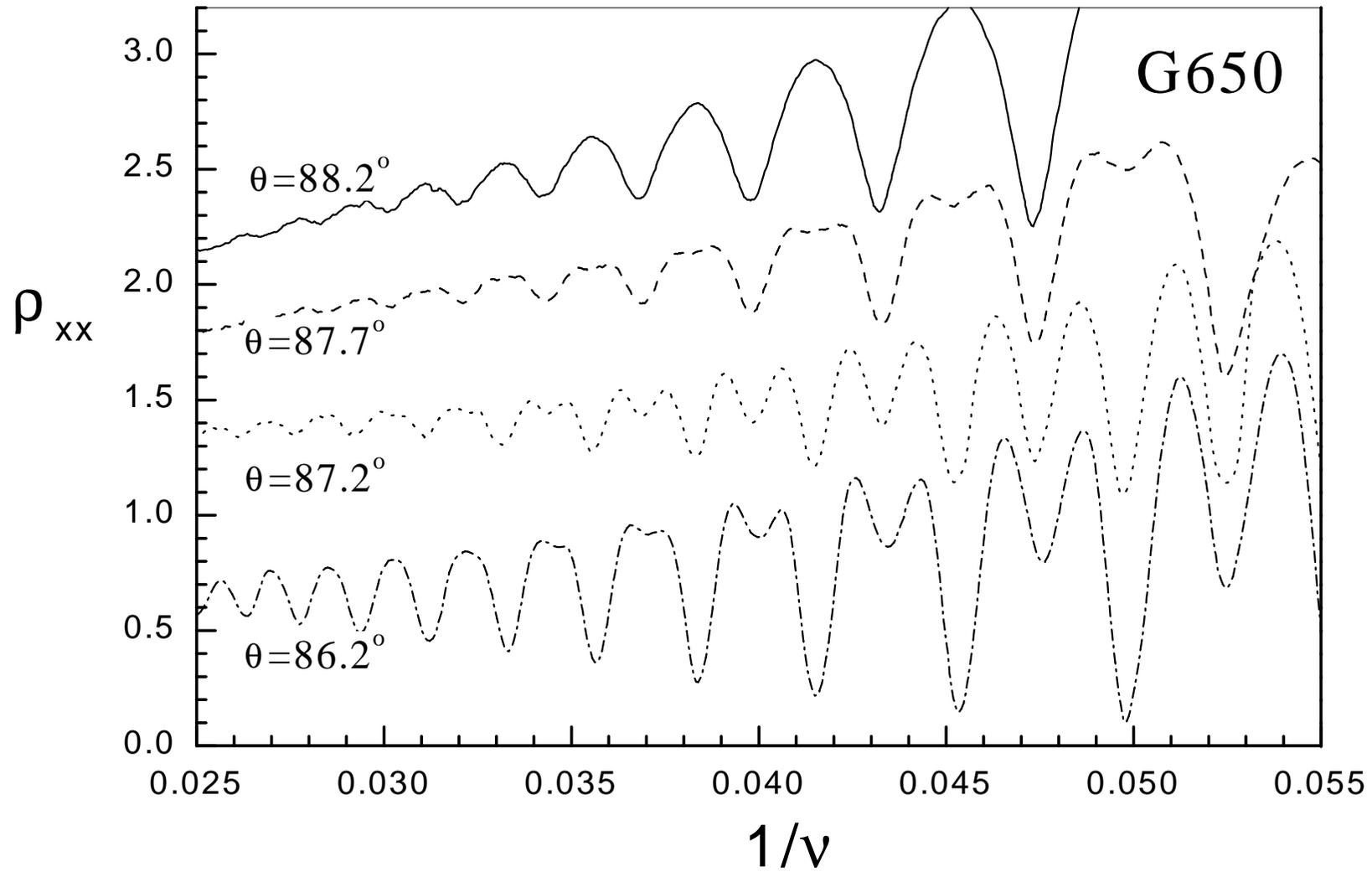
\begin{figure}
\caption{SdHOs at very high tilt angles $\theta$ to the sample normal, showing the swap in 
intensity between even and odd integer $\nu$ and their coincidence at $\theta=87.3^o$.}
\label{fig:coincidence}
\end{figure}

It is well known that for a 2DEG the cyclotron motion, and hence the energy $\hbar\omega_c$, 
and the magnetic length, only depend on the perpendicular component of $B$ whereas the 
Zeeman energy depends on the total magnetic field.  This means that on tilting the sample (such 
that the angle between the magnetic field and a direction normal to the 2DEG is $\theta$) the 
ratio of the Zeeman to cyclotron energy will increase.  When the Zeeman energy is exactly half 
of the cyclotron energy, at some angle $\theta_c$, a ladder of equally spaced levels is generated 
and we would expect that the odd and even integer minima in the SdHO have the same depth.  
At this point we have the condition that $\hbar eB\cos \theta_c/m^*=2\Delta_{spin}$ which 
forms the basis for the coincidence method of determining the $g$-factor \cite{nich}. On further 
tilting other coincidence conditions occur as the levels cross each other.  An example of the first 
condition can be seen in Fig.~\ref{fig:coincidence} for sample G650 at very high tilt angles.  In 
the two lower traces the even integer minima are deeper and on further rotation the upper two 
traces show that the odd integer (spin split) minima dominate, allowing us to deduce that the 
coincidence angle is close to $\theta_c=87.3^o$.  Notice that this is very close to $90^o$ where 
the magnetic field is parallel to the 2DEG and $\cos\theta$ is changing very rapidly.  In the past 
the enhanced $g$-factor has then been extracted by writing $\Delta_{spin}=g^*\mu_BB$ which 
gives the coincidence condition as
\begin{equation}
\cos\theta_c=g^*\frac{m^*}{m_0}
\label{eqn:coin}
\end{equation}
and in this case a value of $g^*=0.71$.  Clearly this is at variance with the value of $g^*=6.8$ 
found in Section~IV.  The same results have also been found in all previous coincidence 
measurements and various explanations have been given, such as there being less exchange 
enhancement at high tilt angle or variable degrees of enhancement depending on the occupancies 
used at the coincidence.

However Eq.~(\ref{eqn:coin}) is not correct, because $\Delta_{spin}$ does not increase linearly 
with the total field.  Although the bare Zeeman part of $\Delta_{spin}$ will increase with total 
field, for an ideal 2DEG the exchange contribution depends only on the perpendicular 
component so we should write 
\begin{equation}
\Delta_{spin}=g_0\mu_BB_{\scriptscriptstyle TOTAL}+\alpha\hbar e B_{\perp}/m^*
\label{eqn:tilt}
\end{equation}
which leads to a new coincidence condition
\begin{equation}
\cos\theta_c=\frac{g_0}{1-2\alpha}\frac{m^*}{m_0}.
\label{eqn:coin2}
\end{equation}
Using a value of $\alpha=0.20$ appropriate for G650 with the coincidence at $87.3^o$ gives a 
value of $g_0=0.4$, in good agreement with the bulk band edge value of 0.44.  In practice we 
would expect a small reduction in the $g$-factor in the 2-D layer due to the effects of 
non-parabolicity thus giving even better agreement.  Our conclusion therefore is that when 
analysed in a way which includes the two dimensional nature of the exchange enhancement the 
coincidence method gives a correct picture of the spin splitting which agrees with other 
measurements.

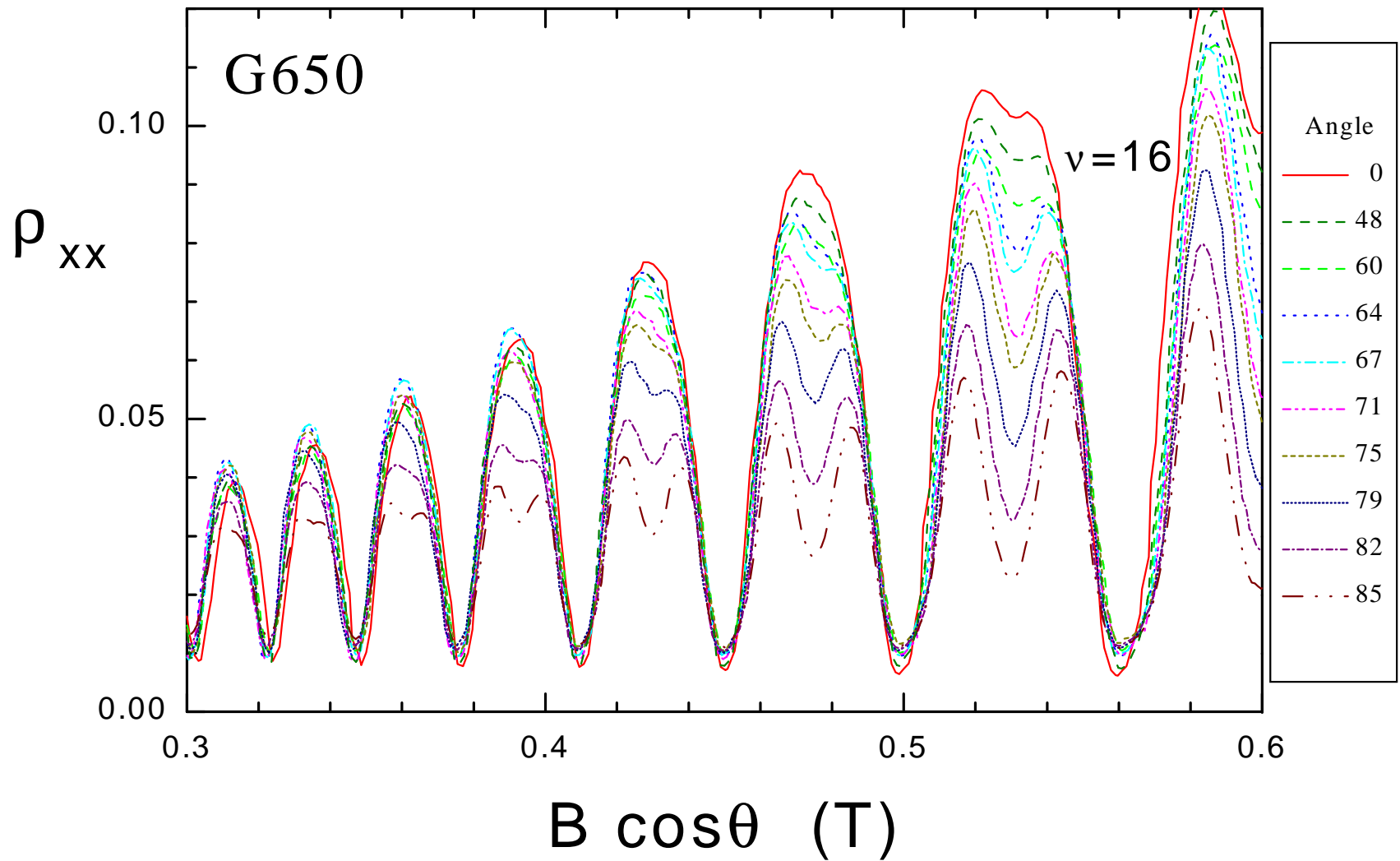
\begin{figure}
\caption{SdHOs for sample G650 in tilted fields at the angles indicated.  Note how the spin split 
minima are deeper and $\nu_c$ increases at higher tilt angles.}
\label{fig:tilt}
\end{figure}

To further substantiate the claim that only the bare Zeeman term increases on tilting, we have 
measured $\nu_c$ as a function of tilt angle.  The resistivity of sample G650 is shown in 
Fig.~\ref{fig:tilt} for angles between 0 and $85^o$ as a function of the normal component of 
magnetic field.  This demonstrates that more spin split peaks are observed at higher tilt angles as 
expected for any mechanism that increases the spin splitting.  By following the resistivity at 
fixed filling factor, it can also be seen that $\delta\nu$ for the spin split maxima displays a very 
similar behavior as a function of tilt to that found earlier as a function of temperature.  In the 
tilting case, the additional parallel field should be considered to open up the spin gap by 
increasing the bare Zeeman contribution, thus delaying the disorder driven collapse just as 
increasing temperature hastened this collapse by adding to the spin disorder.  In Ref.\ 
\cite{wong} the critical density was found to decrease at higher tilt angles which is equivalent to 
saying that the point of collapse moves to lower magnetic fields and is entirely consistent with 
our results.

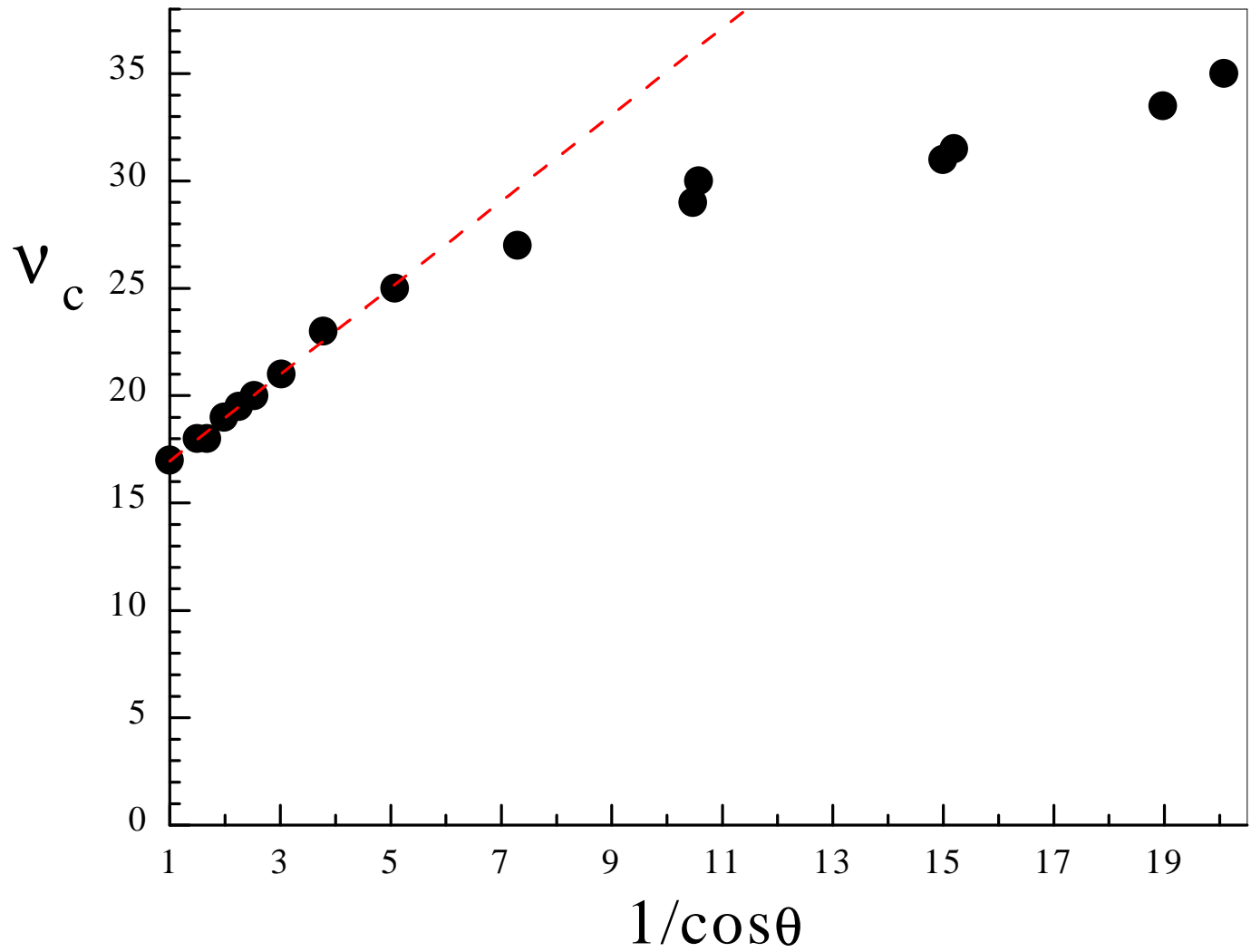
\begin{figure}
\caption{Variation of the filling factor at which spin splitting collapses as a function of tilt angle 
for sample G650.}
\label{fig:tiltnu}
\end{figure}

When only the bare Zeeman energy increases on tilting, the critical filling factor is given by
\begin{equation}
\nu_c(\theta)=n_e\tau_s\ h\left(\frac{g_0}{2m_0\cos\theta }+\frac{\alpha}{m^*} \right)
\label{eqn:nuctheta}
\end{equation}
where it can seen that the increase in $\nu_c$ with angle is only due to the first term -- the 
contribution from the bare Zeeman energy.  The measured critical filling factor is shown as a 
function of tilt in Fig.~\ref{fig:tiltnu} for sample G650.  The line drawn on this figure has the 
gradient predicted by Eq.~(\ref{eqn:nuctheta}), using $g_0=0.40$, with an intercept for $\nu_c$ 
slightly lower than deduced in section V due to the finite temperature of the measurement.  The 
agreement between the theory and experiment clearly demonstrates that only one part of the spin 
splitting is increasing as the field is tilted, as otherwise the gradient would depend on $g^*$ and 
be some 15 times larger.  At the highest angles the data begins to fall below the line. The most 
likely cause of this is a reduction in $\tau _{s}$ caused by the parallel field which will push the 
wavefuction closer to the interface increasing the scattering.  For angles above $85^{o}$ there is 
also a large positive magnetoresistance, suggesting a change in scattering.  However, we were 
not able to verify this directly by measuring $\tau _{s}$ at high tilt angles, precisely because the 
SdHOs are spin split to much higher filling factors.

Several papers have previously measured activation energies in tilted fields in an attempt to 
extract the enhanced $g$-factor from the rate of change of energy gap with tilt angle. The new 
analysis presented above suggests however that accurate and consistent results can only be 
achieved by including the two dimensional nature of the exchange interactions.

\section{Conclusion}

In this paper we have considered how the spin splitting is increased by the addition of an 
exchange term to the bare Zeeman energy.  This enhancement is only present when the spin gap 
is larger than the disorder potential in the sample which can be parameterized by $\tau_s$.  Once 
the exchange enhancement can no longer be sustained the spin splitting collapses critically, 
which is seen both by the separation in filling factor $\delta\nu$ of the spin split maxima and the 
energy gap deduced from the depth of the minima.  We have examined the filling factor $\nu_c$ 
at which this critical collapse occurs as a function of temperature.  Increasing the temperature 
leads to a lower value of $\nu_c$ because thermally excited spins essentially add to the disorder 
potential that the exchange energy must exceed.  A scaling behavior is found which maps the 
temperature dependence of $\delta\nu$ for all filling factors in all the samples on to a single 
curve.  A reliable method of extracting the critical filling factor at $T=0$ from finite temperature 
data is established.  By investigating the variation of $\nu_c$ at zero temperature between 
samples we find a universal empirical relationship $\nu_c = g^*n_e\tau_s h/2m_0$ which is 
completely consistent with the picture of disorder driven collapse.  In tilted magnetic field 
experiments, $\nu_c$ increases as the component parallel to the 2DEG is increased.  This is 
found to be due to only the bare Zeeman component of the spin splitting increasing, not the 
exchange term.  This finding allows us to understand past discrepancies between different 
methods of measuring the enhanced $g$-factor.

\section*{Acknowledgements}
We would like to thank H. Aoki, G.Kido, D.M. Symons and S. Uji for assistance at NRIM, Tsukuba, in making the tilted field measurements.

\newpage

\begin{table}\begin{center}
\caption{Sample parameters (after illumination, except where marked *): spacer thickness 
$L_z$; electron density $n_e$ and mobility $\mu$ at 1~$K$; quantum lifetime $\tau_s$ 
deduced from Eq.~(2).  $g^*$ and $\nu_c$ are discussed in the text.}
\label{samples}

\begin{tabular}{lcddddd}

Sample & $L_z$ & $n_e$ & $\mu$ & $\tau_s$ & $g^*$ & $\nu_c$ \\
 & \AA & $\times10^{15}m^{-2}$ & m$^2$/Vs & ps &  &  \\ \hline

G647 & 3200 & 0.45 & 245 & 6.7 & 8.5 & 11  \\ 
G646 & 2400 & 0.63 & 200 & 10.5 & 7.3 & 21  \\ 
G641 & 1600 & 0.90 & 400 & 13.0 & 7.6 & 31  \\ 
G137 & 1600 & 0.99 & 100 & 4.0 & 6.3 & 15  \\ 
G640 & 1200 & 1.2 & 680 & 7.3 & 6.5 & 23  \\ 
G650 & 400 & 2.2 & 630 & 6.3 & 6.2 & 29  \\ 
G627* & 400 & 2.0 & 310 & - & - & 17  \\ 
G627 & 400 & 3.0 & 370 & - & - & 35  \\ 
G902* & 200 & 3.1 & 148 & 1.1 & 5.1 & 11  \\ 
G902 & 200 & 4.5 & 200 & 2.0 & 4.2 & 20  \\ 
G590* & 100 & 3.2 & 12 & 0.7 & 4.0 & 5  \\ 
G590 & 100 & 6.1 & 65 & 0.9 & 3.4 & 10  \\ 
G378 & 100 & 3.4 & 6 & 0.4 & 3.4 & 3  

\end{tabular}
\end{center}
\end{table}

\begin{references}
\bibitem{N} We will usually refer to the filling factor rather than the LL index.  They are related 
by $\nu=2N+1$ so $\nu=1$ is the spin split minimum of the $N=0$ LL etc.
\end{references}
\end{document}